\documentclass{article}
\usepackage{authblk}

\usepackage[T1]{fontenc}
\usepackage[latin9]{inputenc}
\usepackage{geometry}
\usepackage{xcolor}
\usepackage{pdfcolmk}
\usepackage{float}
\usepackage{amsmath}
\usepackage{amssymb}
\usepackage{graphicx}
\PassOptionsToPackage{normalem}{ulem}
\usepackage{ulem}

\makeatletter

\providecommand{\tabularnewline}{\\}
\floatstyle{ruled}
\newfloat{algorithm}{tbp}{loa}
\providecommand{\algorithmname}{Algorithm}
\floatname{algorithm}{\protect\algorithmname}
\providecolor{lyxadded}{rgb}{0,0,1}
\providecolor{lyxdeleted}{rgb}{1,0,0}

\usepackage{algorithmicx}
\usepackage{setspace}

\@ifundefined{showcaptionsetup}{}{%
 \PassOptionsToPackage{caption=false}{subfig}}
\usepackage{subfig}
\makeatother

\begin{document}

\title{A Point Process Model for Generating Biofilms with Realistic Microstructure and Rheology}
\author[]{%
  J.  A. STOTSKY,
  V.  DUKIC, and  D.  M. BORTZ
}

\affil{%
   Department of Applied Mathematics \\
   University of Colorado \\
   Boulder CO, 80309-0526 \\
   email\textup{\nocorr: \texttt{dmbortz@colorado.edu}}}

\maketitle

\begin{abstract}
Biofilms are communities of bacteria that exhibit a multitude of multiscale
biomechanical behaviors. Recent experimental advances have lead to
characterizations of these behaviors in terms of measurements of the
viscoelastic moduli of biofilms grown in bioreactors and the fracture
and fragmentation properties of biofilms. These properties are macroscale
features of biofilms; however, a previous work by our group has shown that heterogeneous
microscale features are critical in predicting biofilm rheology.
In this paper we use tools from statistical physics to develop a generative
statistical model of the positions of bacteria in biofilms. We show
through simulation that the macroscopic mechanical properties of biofilms
depend on the choice of microscale spatial model. Our key finding
is that a biologically inspired model of the locations of bacteria
in a biofilm is critical to the simulation of biofilms with realistic
\emph{in silico }mechanical properties and statistical characteristics.
\end{abstract}

\section{Introduction}

Biofilms are complex, multi-organism communities of bacteria. They
are abundant in nature and grow readily in many industrial systems
where they often cause maintainence and safety issues. Measures to
mitigate or remove biofilms, though costly, are often necessary in
the design and operation of many industrial systems \cite{flemming2011microbial,guelon2011advances}.
  The demand for better biofilm control strategies and the causative
role of biofilms in bacterial infections has inspired the development
of numerous mathematical biofilm models \cite{gaboriaud_coupled_2008,guelon2011advances,hammond2013viscosity,laspidou2004modeling,stotsky2015rheology,sudarsan2015simulating,vo2010experimentally,zhang2008phase1,zhang2008phase2,zhao20163d}.
 These models have been designed to capture a wide range of biofilm-related
phenomena such as the biomechanical response to mechanical forces,
growth dynamics, and persistence in the presence of antimicrobial
substances. Yet, despite a thirty year history, it is only recently
that the validation of models by comparison with empirical data has
become a focus in computational biofilm studies. Recent works have
shown agreement between experiment and simulation regarding the frequency
dependent dynamic moduli \cite{stotsky2015rheology}, the ranges of
certain spring constants \cite{vo2010experimentally}, and the rate
of disinfection of biofilms in response to antimicrobial substances
\cite{zhao20163d}. However, this recent progress has also led to
new questions.

Due to the complex behavior of biofilms, even state-of-the-art
models rely on many assumptions.
Material parameter values, biofilm morphology, and connectivity of biofilm networks are often
specified heuristically, and even when parameters are informed by
experimental results, there are difficulties. For instance, the experimental
measurements of biofilm properties may differ by orders of magnitude
between studies depending on the specifics of the conditions under
which biofilms were grown and tested \cite{pavlovsky2013situ}. 

The assumptions noted
before involve microstructural properties. Although, some microstructural
properties have been measured \cite{dzul2011contribution,stewart2013role},
little work has been done to elucidate the influence of microstructural
properties on macroscale behaviors. Along these lines, there are two
main goals to this paper. The first is to develop a microstructural
description of \emph{S. epidermidis} biofilms by estimating certain
fundamental statistical characteristics from experimental data. The
second is to numerically demonstrate the efficacy of first and second
order summary statistics for generating data with similar material properties
as experimental data sets. The similarity of material properties is
tested through simulation using the \emph{heterogeneous rheology Immersed Boundary Method} (hrIBM) \cite{hammond2013viscosity,stotsky2015rheology}.

This paper is organized as follows. In Section \ref{sec:Statistics-of-Biofilm},
we provide an overview of some non-parametric estimators of summary
statistics of finite point processes. These estimators are based on those discussed in 
\cite{stoyan_estimation_1993,stoyanstochastic,moller2003statistical,kerscher2000comparison}. We then apply these estimators
to four data sets to compute the intensity, pair correlation function,
and nearest neighbor distributions of each data set. The experimental
data sets, obtained using the techniques described in \cite{stewart2013role,pavlovsky2013situ},
consist of three dimensional coordinates of the centers of mass of
approximately $4000$ bacteria from live biofilms. 

In Section \ref{sec:Replication-of-Biofilm}, we introduce a statistical
model for the arrangement of bacteria in a biofilm parametrized by the summary
statistics discussed in Section \ref{sec:Statistics-of-Biofilm}.
The model relies on results from the statistical physics of fluids
and is designed to accurately replicate first and second order interactions
between bacteria. From the model, an unnormalized probability density associated with the configuration of bacteria in the biofilm is obtained and used in Markov Chain Monte Carlo (MCMC) simulations
to generate ``artificial'' biofilms \cite{moller2003statistical,yeong1998reconstructing}.
Unnormalized probability density functions are used due to the difficulty in computing
the normalization constant needed to obtain a probability density of configurations.
MCMC algorithms avoid dependence on normalizing constants since they
only require ratios of probability densities. 

In Section \ref{sec:Comparison-of-Material}, we compare the material
properties of the artificial biofilms generated by the statistical
model presented in this paper, along with some previous biofilm generation
models (e.g. \cite{alpkvist2007description,sudarsan2015simulating}),
to data obtained through high resolution microscopy techniques. This
comparison is achieved through simulations using the hrIBM
to estimate the dynamic moduli of the resulting biofilms. In Section
\ref{sec:Discussion}, we discuss the results, motivate some future
research directions, and suggest potential improvements. 

\section{Summary Statistics of Point Data\label{sec:Statistics-of-Biofilm}}

In the spatial statistics literature, summary statistics such as the
\emph{number density} (commonly called the \emph{intensity}), \emph{pair
correlation function}, and \emph{nearest neighbor distributions} are
frequently used to analyze random data. These summary statistics first
appeared in diverse disciplines such as geography, forestry, and statistical
physics; but have become more unified in recent years as the field
of spatial statistics has evolved. Two dimensional estimates for these
quantities are abundant \cite{stoyan_estimation_1993,van2011aj,baddeley2000non,guan2008consistent,
lieshout1996nonparametric,guan2007least,kolaczyk_nonparametric_2000,
ripley1991statistical,stoyanstochastic},
and three dimensional estimators, which are the focus of this work,
have been applied to problems in astronomy and cosmology \cite{davis1983survey,kerscher2000comparison}. In our discussion, we follow the
notation of \cite{stoyanstochastic} and \cite{moller2003statistical}
where concise introductions to many of the quantities discussed here
can be found. In Table \ref{tab:Definitions-of-commonly} the most
frequently used symbols are defined. In Appendices \ref{subsec:Estimation-Density}
and \ref{subsec:Estimation-PCF}, we adapt optimization strategies
from non-parametric density estimation to our estimates of the number
density and pair correlation function.

\begin{table}
\caption{\label{tab:Definitions-of-commonly}Definitions of commonly used symbols.
The $n$th order probability density when integrated over a product
set, $B_{1}\times B_{2}\times\dots B_{n}$ gives the probability that
for each $\boldsymbol{r}_{i}$, there exists a $B_{k}$ such that
$\boldsymbol{r}_{i}\in B_{k}$ for $i\in[1,n]$.}
\begin{centering}
\begin{tabular}{c c}
\hline 
Symbol & Definition\tabularnewline
\hline 
\hline 
$W$ &  Borel set containing experimental data\tabularnewline
\hline 
$\boldsymbol{r}$, $(x,\,y,\,z)$ & point in $\mathbb{R}^{3}$\tabularnewline
\hline 
$\boldsymbol{r}^{n}$ & collection of $n$ points in $\mathbb{R}^{3}$\tabularnewline
\hline 
$\Phi$ & realization of a point process\tabularnewline
\hline 
$\Phi(B)$ & number of points of $\Phi$ in set $B$\tabularnewline
\hline 
$\nu(B)$ & Lebesgue measure of a set, $B$\tabularnewline
\hline 
$\rho^{(2)}(\boldsymbol{r}_{1},\boldsymbol{r}_{2})$ & 2nd order factorial moment density\tabularnewline
\hline 
$g^{(2)}(\boldsymbol{r}_{1},\boldsymbol{r}_{2})$ & 2nd order correlation function\tabularnewline
\hline 
$h^{(2)}(\boldsymbol{r}_{1},\boldsymbol{r}_{2})$ & indirect correlation function, $h^{(2)}=g^{(2)}-1$\tabularnewline
\hline 
$f^{(n)}(\boldsymbol{r}^{n})$ & $n$th order probability density\tabularnewline
\hline 
$\rho(\boldsymbol{r})$ & number density (also known as the intensity), equivalent to $\rho^{(1)}(\boldsymbol{r})$\tabularnewline
\hline 
$\hat{q}$ & estimator for quantity $q$\tabularnewline
\hline 
$\tilde{q}$ & Hankel transform of quantity $q$\tabularnewline
\hline 
$\mathbb{E}[q]$ & expectation of $q$\tabularnewline
\hline 
$k_{b}(r)$ & smoothing kernel with support of radius $b$\tabularnewline
\hline 
$c_{b}(W)$ & edge correction factor\tabularnewline
\hline 
$c(\boldsymbol{r}_{1},\boldsymbol{r}_{2})$ & direct correlation function\tabularnewline
\hline 
$\phi(\boldsymbol{r})$  & singlet potential\tabularnewline
\hline 
$v(\boldsymbol{r}_{1},\boldsymbol{r}_{2})$ & pair potential\tabularnewline
\hline 
$\beta$ & temperature\tabularnewline
\hline 
$\boldsymbol{e}_{x}$ & unit vector in $x$-coordinate direction\tabularnewline
\hline 
$H_{P}(r)$ & nearest-neighbor distribution function\tabularnewline
\hline 
$G^{\prime},\ G^{\prime\prime}$ & first and second dynamic moduli defined as  $G^{\prime}=(\sigma/\epsilon)\cos\delta$\tabularnewline
& and $G^{\prime\prime}=(\sigma/\epsilon)\sin\delta$, where  $\sigma$ is a stress amplitude,\tabularnewline
&  $\epsilon$ a strain amplitude, and $\delta$ a phase lag (see \cite{pavlovsky2013situ}) \tabularnewline
\hline
\end{tabular}
\par\end{centering}
\end{table}
Each realization of a point process is a random, countable set of
distinct points in $\mathbb{R}^{3}$, denoted by $\Phi$. The restriction,
$\Phi_{W}\equiv\Phi\cap W$, of the point process to a Borel set,
$W$, containing the experimental data is of practical significance
since experimental data is always contained in a bounded region. This
restriction can lead to theoretical and numerical difficulties, and
edge correction factors must often be derived in order to obtain accurate
estimators of the statistical quantities of interest \cite{stoyan_estimation_1993,ripley1991statistical,jones1993simple}.
As will be shown in Sections \ref{subsec:Point-Process-Number}, \ref{subsec:PairCorrelation},
\ref{subsec:Nearest-Neighbor-Distributions}, each statistic we compute
from the data generally requires its own edge correction factor. 

\begin{figure}
\begin{centering}
\includegraphics[width=0.7\columnwidth]{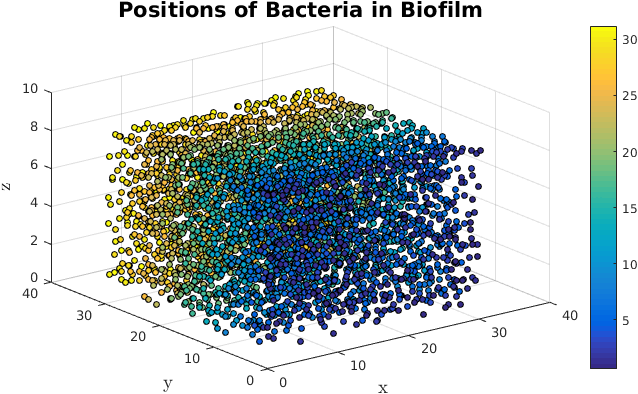}
\par\end{centering}
\caption{\label{fig:An-experimental-data-set}An experimental data set from
a live biofilm. The data set is approximately $30\mu m\times30\mu m\times10\mu m$
in size and consists of the centers of mass of 3981 live bacteria.
This data was obtained from high-resolution confocal microscopy images.
The techniques used to obtain this data are discussed in \cite{stewart2013role}.
The coloring is for ease of viewing and simply corresponds to the
$y$ coordinate of each bacterium.}
\end{figure}

The experimental data available to us consists of sets of points in
three dimensional space. Each point in a data set corresponds to the
center of mass of an individual bacterium in a live biofilm. In Figure
\ref{fig:An-experimental-data-set}, we depict one of the four experimental
data sets available to us. As indicated by the axes, the $z$ coordinate of each data
point will indicate its vertical location, and the $x$ and $y$ coordinates
will indicate the horizontal location and depth of each point. 

Although the complete statistical characterization of a point process is impractical, previous
statistical studies show that information garnered from low order
statistics of point processes is often sufficient to understand key
features of data \cite{baddeley2000non,baddeley2000practical,lieshout1996nonparametric,moller2003statistical,stoyanstochastic}.
When plausible, models based on first and second order statistics
are preferred since higher order statistics are difficult to estimate
when only limited amounts of data are available and often are computationally
expensive to compute when data is abundant. Thus, we proceed by estimating
first and second order statistics of the biofilm data. In Section
\ref{sec:Comparison-of-Material}, we qualitatively validate the ability
of first and second order statistics to characterize the arrangement
of bacteria in \emph{S. epidermidis} biofilms. 

\subsection{Point Process Number Density\label{subsec:Point-Process-Number}}

To characterize the first order properties of a spatial point process,
we ought to be able to estimate the average number of points in
various subsets of the domain containing the data. In many cases,
this average number of points is assumed simply to be proportional
to the volume of the subset in consideration. However, a more accurate
accounting of the first order properties of a points process requires
the possibility of spatial variation. 

For a general point process, under mild restrictions (see \cite[\S 9.2]{daley2007introduction}),
the average number of points in some region, $B$, can be written
as the Lebesgue integral of a function (known as a \emph{rate}), $\rho(\boldsymbol{r})$,
\[
\mathbb{E}\left[\Phi(B)\right]=\int_{B}\rho(\boldsymbol{r})d\boldsymbol{r}.
\]
This function, known as the \emph{number density} or \emph{intensity},
is of fundamental interest in the analysis of point processes. When
presented with empirical point data, a typical goal is to infer
whether the number density is constant, signifying a \emph{stationary }point
process, or spatially variable, indicating a \emph{non-stationary}
point process. To study the stationarity of a point process, an estimation
procedure must be implemented to approximate the number density. When
intuition about the nature of possible intensity variations is lacking,
non-parametric estimators are often the tool of choice for the estimation
of the number density. Since we are not aware of any previous models
of the spatial statistics of bacteria in biofilms, a non-parametric density
estimation is used in this work to allow for as much generality as
possible.

In addition to being flexible, non-parametric
estimates of the number density are often easy to implement and versatile,
yielding suitable intensity approximations for many classes of point
processes. However, a common source of concern with non-parametric
estimates is the presence of statistical bias \cite{rosenblatt1956remarks,wand1993comparison,hall1991local}.
Non-parametric estimates are rarely pointwise unbiased\footnote{This is similar in principle to non-parametric density estimation.
A proof of the non-existence of pointwise unbiased estimators for
probability density functions is given in \cite{rosenblatt1956remarks}}, but are often asymptotically unbiased. Heuristically, the bias is
due to inaccuracy in the numerical approximation of a Radon-Nikodym
derivative \cite{billingsley2008probability} and is analogous to
the truncation error associated with finite difference approximations
of differentiable functions. Furthermore, as finite difference approximations
depend on a spatial increment, non-parametric number density estimators
are parametrized\footnote{The term\emph{ non-parametric }refers to the absence of an underlying
 assumption about the functional form of $\rho(\boldsymbol{r})$,
not the absence of any tuning parameters in the estimator. Aside from
a few simple examples, non-parametric estimators almost always have
some sort of parameter which must be chosen based on the scales of
variation seen in the data. } by a scale parameter, $b$. Smaller values of $b$ reduce the bias,
but increase the variance of the estimator \cite{parzen1962estimation,rosenblatt1956remarks,wand1993comparison}.
To ensure efficient estimators, careful selection of $b$ is needed
to balance this tradeoff between bias and variance. Further discussion of this balancing of bias and variance is included in the appendix. 

\subsubsection{Intensity Estimators}
\begin{figure}
\begin{centering}
\subfloat[\label{fig:Density_Data}]{\begin{centering}
\includegraphics[width=0.35\columnwidth]{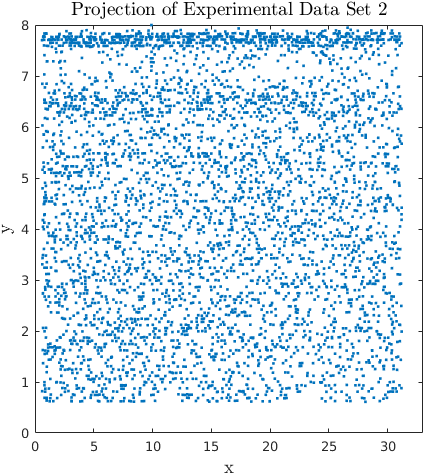}
\par\end{centering}
}\subfloat[\label{fig:Density_Graphs}]{\begin{centering}
\includegraphics[width=0.35\columnwidth]{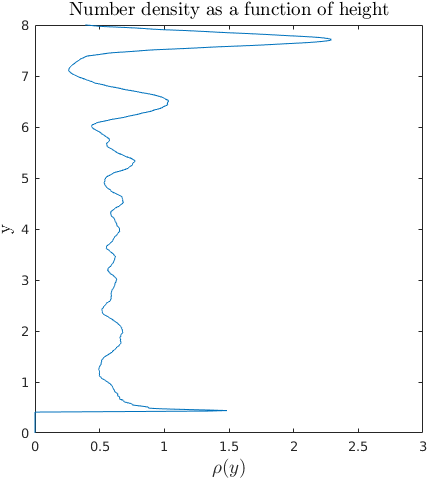}
\par\end{centering}
}
\par\end{centering}
\caption{\label{fig:The-height-dependent-intensity}\ref{fig:Density_Data}:
The projection of one of the data sets into the $xz$-plane is shown.
The number density variation near the top of the domain can clearly be observed
and some oscillatory intensity variation is discernible. \ref{fig:Density_Graphs}:
The height dependent intensity, $\rho(y)$, from the same experimental
data set is depicted.  We used a scale parameter, $b=0.2$
for all four data sets.}
\end{figure}

Our first approach was to examine the dependence of the intensity on the $x$, $y$, $z$ individually (for instance, assume that $\rho(\boldsymbol{r})=\rho(x)$, independent of $y$ and $z$). With this approach, we  
found that although the bacteria locations are three dimensional coordinates,
the number density is observed to vary only along the $z$-axis, with
changes in the distance from the fluid-biofilm interface. As depicted in Figure \ref{fig:Density_Data},
it appears as though bacteria near the fluid-biofilm interface stratify
into layers and pack more closely together than cells further interior.
In Figure \ref{fig:Density_Graphs}, this phenomenon is clearly seen as bumps in the number density.  To
quantify this variation, we use an estimator based on the discussion
in \cite[\S 4]{stoyanstochastic}. Given a realization of a point
process, the height dependent intensity is estimated by
\begin{equation}
\hat{\rho}(z)=\sum_{\boldsymbol{r}_{i}\in\Phi}\frac{k_{b}(\boldsymbol{r}_{i}\cdot\hat{\boldsymbol{e}}_{z}-z)}{c_{b,W}(z)}\label{eq:DensityVerticalEstimator}
\end{equation}
where $k_{b}(z)=\left(3/(4b)\right)\left(1-\left(z/b\right)^{2}\right)\mathbf{1}_{|z|<b}$,
is the \emph{Epanechnikov kernel} \cite{epanechnikov1969non}, a commonly
used density estimation kernel, with scale parameter, $b$. For notation, we use a hat (i.e. $\hat{\rho}$) to denote an estimator for a quantity. 
Defining $A$ as the area of a cross-section perpendicular to the $z$-axis and $W$ to be the domain containing the data,
the denominator, $c_{b,W}(z)$ in Equation \eqref{eq:DensityVerticalEstimator},
is an edge correction factor defined as 
\[
c_{b,W}(z)=A\int_{[z-b,z+b]\cap W}k_{b}(z^{\prime}+z)dz^{\prime}.
\]

The number density estimates obtained from data sets \#1-4 exhibit some variation
near the biofilm-fluid interface. Since the number of realizations
(in our case, 4) is insufficient to propose a random intensity model,
such as a Cox process model \cite[\S 6.2]{daley2007introduction},
we parametrize our model on the intensity from a particular data set in
each simulation in Section \ref{sec:Replication-of-Biofilm}. 

We also found that the precise form of the number density variation near the top of the biofilm did not have a substantial impact on the overall material properties of the biofilm. This was tested by conducting simulations in which the area of number density variation was removed from the data and observing that the difference in the strength of the biofilm (see Section \ref{sec:Comparison-of-Material}) was not strongly effected.

As previously mentioned, the value of the scale parameter, $b$, has
a significant impact on the resulting number density estimate and should
be carefully chosen to balance variance and bias. If $b$ is too small,
the estimate will be noisy, whereas if $b$ is too large, key features
of the data will be blurred. The estimation of a one dimensional intensity
function is quite similar to the non-parametric estimation of a probability
density up to the normalization condition required of probability
densities. This motivates the adaptation of the \emph{Least Squares
Cross-Validation} (LSCV) technique, a common optimization strategy
for non-parametric probability density estimators \cite{wand1993comparison},
to optimize our choice for $b$ in Equation (\ref{eq:DensityVerticalEstimator}).
We refer the interested reader to Appendix \ref{subsec:Estimation-Density}
for a discussion of LSCV bandwidth selection. 

In addition to the number density, we will see in Section \ref{subsec:Computation-of-the-Potentials}
that the gradient of the number density is also needed in the computation
of a spatially dependent potential energy function. In order to compute
this quantity, we adhere to the strategy discussed in \cite{wand1993comparison}
and use the differentiable, triweight kernel density estimator, defined
as
\[
T_{b}(z)=\frac{35}{32b}(1-(z/b)^{2})^{3}\boldsymbol{1}_{\{|z|\leq b]}.
\]
Similar to the Epanechnikov kernel, the triweight kernel is a symmetric
probability density function. However, it is more useful for density
derivative estimation since it is twice differentiable whereas the
Epanechnikov kernel is not differentiable. The number density derivative
is approximated by
\[
\widehat{\rho_{z}}(z;b)=\frac{1}{A}\sum_{\boldsymbol{r}_{i}\in\Phi\cap W}\frac{\partial T_{b}(z-z_{i})}{\partial z}.
\]
As derivatives are more sensitive to noisy data, optimal balancing
of variance and bias in number density derivative estimation tends to yield a larger
value of $b$ than that used in number density estimation \cite{hardle1990bandwidth}. By inspection of the resulting estimators, we found that a value of $b=0.5$ seemed to give the best result. To correct for bias near the boundaries of the domain,
adjustments to the kernel constructed by the technique in \cite[Equation (8.2)]{jones1993simple}
 were used.  

\subsection{Pair Density and Pair Correlation Functions\label{subsec:PairCorrelation}}

The second order interactions of a spatial point process with nonzero
number density can be characterized through either the
\emph{pair correlation function}(PCF) or the \emph{second order factorial
moment density}(SOFMD). Both functions measure the tendency of two
points in space to be jointly contained in a realization of the point process.
For two arbitrary points, $\boldsymbol{r}_{1}$ and $\boldsymbol{r}_{2}$,
the SOFMD, denoted by $\rho^{(2)}(\boldsymbol{r}_{1},\boldsymbol{r}_{2})$
is defined for any two sets, $B_{1}$ and $B_{2}$, through the integral
relation
\[
\mathbb{E}[\Phi(B_{1})\times\Phi(B_{2})]=\mathbb{E}\left[\Phi(B_{1}\cap B_{2})\right]+\int_{B_{2}}\int_{B_{1}}\rho^{(2)}(\boldsymbol{r}_{1},\boldsymbol{r}_{2})d\boldsymbol{r}_{1}d\boldsymbol{r}_{2}.
\]
Note that for disjoint sets, the SOFMD is simply the expectation of
the product of the number of points in each set. In fact, $n$th order 
factorial moment densities are defined for disjoint sets as 
\[
\mathbb{E}\left[\Phi(B_{1})\cdot\Phi(B_{2})\cdot\dots\cdot\Phi(B_{n})\right]=\int_{B_{n}}\dots\int_{B_{1}}\rho^{(n)}(\boldsymbol{r}_{1},\dots,\boldsymbol{r}_{n})d\boldsymbol{r}_{1}\dots d\boldsymbol{r}_{n}.
\]
For nondisjoint sets, the formulas for the factorial moment densities
of arbitrary order are more complex and not listed here. However,
there is some discussion in \cite{stoyanstochastic} of how such formulas
may be derived.  

The PCF, denoted
$g^{(2)}(\boldsymbol{r}_{1},\boldsymbol{r}_{2})$, is defined as a
rescaling of $\rho^{(2)}(\boldsymbol{r}_{1},\boldsymbol{r}_{2})$
by the number density, 
\[
g^{(2)}(\boldsymbol{r}_{1},\boldsymbol{r}_{2})=\frac{\rho^{(2)}(\boldsymbol{r}_{1},\boldsymbol{r}_{2})}{\rho(\boldsymbol{r}_{1})\rho(\boldsymbol{r}_{2})}.
\]
For simplicity of notation, the superscript, $(2)$, on $g^{(2)}(\boldsymbol{r}_{1},\boldsymbol{r}_{2})$
will be omitted except where there is ambiguity between $g^{(2)}$
and $g^{(n)}$ for some $n\neq2$.\footnote{General $n$-point
correlation functions that correspond to $n$th order\emph{ factorial
moment densities} can be defined as
\[
g^{(n)}(\boldsymbol{r}_{1},\dots,\,\boldsymbol{r}_{n})=\frac{\rho^{(n)}(\boldsymbol{r}_{1},\dots,\,\boldsymbol{r}_{n})}{\prod_{i=1}^{n}\rho^{(1)}(\boldsymbol{r}_{i})}.
\]}

The dependence of the PCF on its arguments 
can be simplified when the underlying point process is
stationary or isotropic and stationary. For stationary point processes,
the PCF is solely a function of $\boldsymbol{r}_{1}-\boldsymbol{r}_{2}$
and, when the point process is also isotropic, it is further
constrained to be dependent only on the radial distance, $\vert\boldsymbol{r}_{1}-\boldsymbol{r}_{2}\vert$.
When the number density is low enough (but nonzero) and the point
process has a finite interaction radius, the PCF
will tend towards one as $\vert\boldsymbol{r}_{1}-\boldsymbol{r}_{2}\vert\rightarrow\infty$.

\begin{figure}
\centering{}\subfloat[\label{fig:PCF_3d}]{\begin{centering}
\includegraphics[width=0.45\columnwidth]{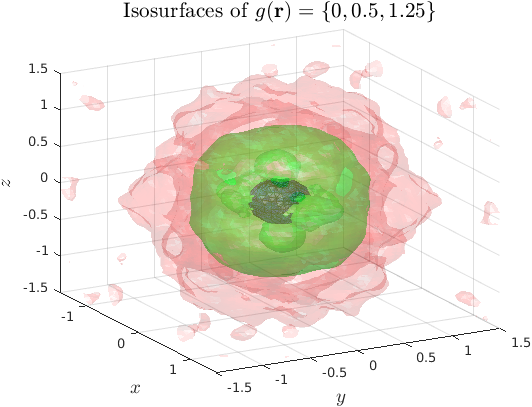}
\par\end{centering}
}\subfloat[\label{fig:PCF_1d}]{\begin{centering}
\includegraphics[width=0.45\columnwidth]{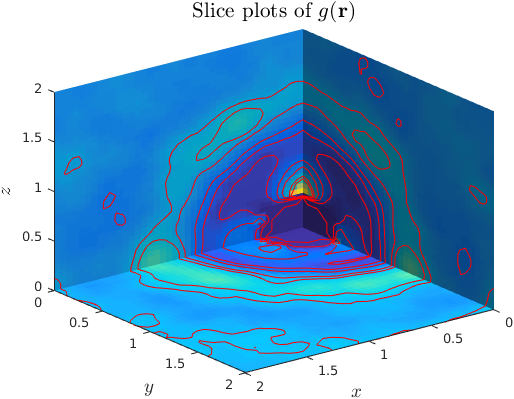}
\par\end{centering}
}\caption{\label{fig:Pair-Correlation-Functions}Both images are of PCFs averaged
over the four experimental data sets. In Figure \ref{fig:PCF_3d}:
isosurfaces of the anistropic pair correlation function at $g^{(2)}(\boldsymbol{r})=0$
(blue), $0.5$ (green), and $1.25$ (red) are depicted, and Figure
\ref{fig:PCF_1d}: contains slice plots and contours of the anisotropic
pair correlation function. The anisotropic PCF is related to the isotropic
PCF through the integral $g_{i}^{(2)}(r)=\frac{1}{4\pi}\int\int g_{a}^{(2)}(r,\phi,\theta)\sin\theta d\theta d\phi.$
The anistropic pair correlation function shows some vertical anistropy,
however most of the variation appears as a radially symmetric function.}
\end{figure}

\subsubsection{Pair Correlation Function Estimates}
We would like to apply these simplifications since we are constrained by the finite size of the data sets. However, neither simplification mentioned above is directly applicable to point processes with variable intensity. 
 In \cite{baddeley2000non}, the idea of a \emph{second order intensity
reweighted stationary} (SOIRS) point process is introduced. This idea is defined (assuming the intensity is nonzero) by considering the random measure
\[
\Xi(B) = \sum_{\boldsymbol{x}_i\in\Phi\cap B}\frac{1}{\rho(\boldsymbol{x}_i)}.
\]
If the second moment of $\Xi$ is stationary, then the point process is deemed to be SOIRS.
The advantage of such an assumption in our application, is that is allows for the approximation of a radially symmetric pair correlation of the form \cite{baddeley2000non,guan2007least}
\begin{equation}
\hat{g}(r)=\frac{1}{4\pi r^{2}\bar{\gamma}_{W}(r)}\sum_{\{\boldsymbol{r}_{i},\boldsymbol{r}_{j}\}\in\Phi\cap W}^{i\neq j}\frac{k_{b}(r-\vert\boldsymbol{r}_{i}-\boldsymbol{r}_{j}\vert)}{\rho(\boldsymbol{r}_{i})\rho(\boldsymbol{r}_{j})},\label{eq:RadiallySymmPCF}
\end{equation}
where $\bar{\gamma}_{W}(r)$ is the isotropized set covariance of $W$ \cite{stoyanstochastic}.

 Such an assumption, if justified, is beneficial since it allows us
to compensate for variable number density while allowing $g(\boldsymbol{r}_{1},\boldsymbol{r}_{2})$
to remain a radially symmetric function. Following \cite{stoyan_estimation_1993},
the expectation of $\hat{g}(r)$ is of the form :
\begin{equation}
\mathbb{E}\left[\hat{g}(r)\right]=\int g(\vert\boldsymbol{r}^{\prime}\vert)k_{b}(\vert\boldsymbol{r}^{\prime}\vert-r)d\boldsymbol{r}^{\prime}.\label{eq:Expectation_PCF}
\end{equation}

Taking the limit as $b\rightarrow0$ in Equation (\ref{eq:Expectation_PCF})
shows that, under a few restraints on $k_{b}(r)$, $\hat{g}(r)\rightarrow g(r)$
at every point of continuity of $g(r)$. Thus, $\hat{g}(r)$ is asymptotically
pointwise unbiased for continuous pair correlation functions. For discontinuous pair correlation functions, such as those arising
from hard-sphere processes, the estimate is not asymptotically pointwise
unbiased at the point of discontinuity, but is asymptotically convergent
to $g(r)$ in the least squares sense.

In computing $\hat{g}(r)$, the estimator in Equation (\ref{eq:RadiallySymmPCF})
is subject to an additional bias when $\rho(\boldsymbol{r})$ is approximated
using the non-parametric estimator in Equation (\ref{eq:DensityVerticalEstimator}).
In \cite{baddeley2000non}, it is shown that this source of bias can
be reduced if the process is a Poisson process, and $\hat{\rho}(\boldsymbol{r})$
is altered when $\boldsymbol{r}\in\Phi$ such that, 
\[
\hat{\rho}(\boldsymbol{r})=\sum_{\boldsymbol{r}_{i}\in\Phi\backslash\{\boldsymbol{r}\}}\frac{k_{b}(\boldsymbol{r}-\boldsymbol{r}_{i})}{c_{b}(\boldsymbol{r})}=-\frac{k_{b}(0)}{c_{b}(\boldsymbol{r})}\boldsymbol{1}_{[\boldsymbol{r}\in\Phi]}+\sum_{\boldsymbol{r}_{i}\in\Phi}\frac{k_{b}(\boldsymbol{r}-\boldsymbol{r}_{i})}{c_{b}(\boldsymbol{r})}.
\]
 For hard-sphere processes, the bias may also be reduced by this adjustment,
but there is no longer a guaranteed improvement. This is because the
bias that results from evaluating $\rho(\boldsymbol{r})$ at points
of $\Phi$ can be positive or negative for hard-sphere processes,
whereas it is always positive for Poisson processes. 

\begin{figure}
\centering{}
\subfloat[\label{fig:PCF_comp}]{\begin{centering}
\includegraphics[width=0.45\columnwidth]{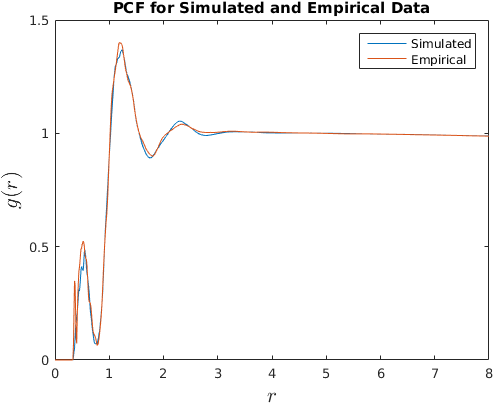}
\par\end{centering}
}\subfloat[\label{fig:Density_comp}]{\begin{centering}
\includegraphics[width=0.45\columnwidth]{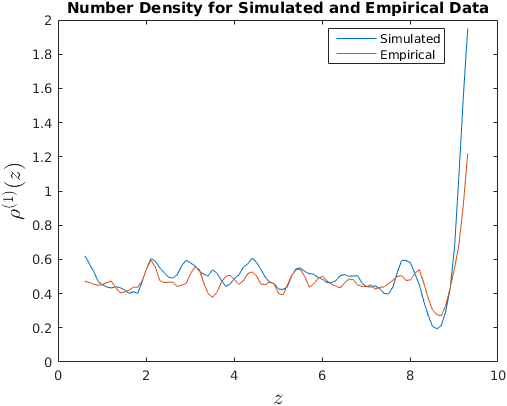}
\par\end{centering}}
\caption{\label{fig:RadiallySymmPCF} \ref{fig:PCF_comp}: Comparison between experimental and simulated data of the SOIRS estimator for a radially symmetric pair correlation function. The radially symmetric pair correlation
exhibits peaks at $r\approx0.6$, $r\approx1.2$, and $r\approx2.4$
micrometers. The relative heights of the first two peaks is an interesting
finding, and has been observed in biofilm data in a previous study
\cite{stewart2013role}. The first, smaller peak is likely due to
a small number of bacteria that were undergoing cellular division
at the time the data set was recorded although the possibility of
some experimental error cannot be definitively ruled out \cite{stewart2015artificial}. \ref{fig:Density_comp}: Comparison of number density between experimental data and simulated data. The simulation method used to generate data is discussed in Section \ref{sec:Replication-of-Biofilm}.}
\end{figure}

In Figures \ref{fig:PCF_3d} and \ref{fig:PCF_1d}, an anisotropic and a radially
symmetric estimate of $g(\boldsymbol{r})$ averaged over four data sets are depicted, and in Figure
\ref{fig:RadiallySymmPCF} the SOIRS radially symmetric PCF estimator
defined in (\ref{eq:RadiallySymmPCF}), computed from one data set, is depicted.

An intriguing property of the biofilm data is the presence of two
prominent peaks in the pair correlation function. It has been suggested that
the first, smaller peak in Figure \ref{fig:Pair-Correlation-Functions}
is indicative of bacteria undergoing cell division at the time their
position was measured \cite{stewart2015artificial}. The second peak occurs near the average diameter
of a non-dividing bacterium. The third peak at $r\approx2.4\mu m$
is is indicative of regularity in the data \cite[\S 5]{hansen1990theory}
and is likely the effect of more distant bacteria.

Recall from Figure \ref{fig:The-height-dependent-intensity} 
that there is evidence of variability in  
the number density along the $z$ axis. This motivated our assumption of a SOIRS process
in analyzing the data. However, we have not provided any evidence against an alternative hypothesis: that the PCF is not translation invariant. 
To make a convincing argument in favor of the SOIRS assumption (or at least that it is reasonable), it is necessary
to estimate the magnitude of the variability of the pair correlation
function. 

To test for variability in the pair correlation function, its estimator, $\hat{g}(r)$
is calculated for various subsets of the full data set, $\Phi\cap W$.
In particular, we compute $\hat{g}(r)$ over sets of the form $V_{z}=[z-\Delta z,\,z+\Delta z]\cap W$
with $z\in(\min(z_{i}),\max(z_{i}))$ where $z_{i}$ are the $z$-coordinates
of the bacteria, and $\Delta z=1\mu m$. Computing $\hat{g}(r)$ on $V_z$ as a function of $z$, we found only small variations in the resulting PCF.
To fully utilize the available data in these computations, we use
an altered estimator of $g(r)$, denoted $\hat{g}_{V_{z}}(r)$, 
\[
\hat{g}_{V_{z}}(r)=\frac{1}{4\pi r^{2}\bar{\gamma}_{V_{z},W}(r)}\sum_{\boldsymbol{r}_{i}\in V_{z}\cap\Phi}\sum_{\boldsymbol{r}_{j}\in W\cap\Phi\backslash\{\boldsymbol{r}_{i}\}}\frac{k_{b}(r-\vert\boldsymbol{r}_{i}-\boldsymbol{r}_{j}\vert)}{\rho(\boldsymbol{r}_i)(\rho(\boldsymbol{r}_j)}
\]
 with 
\[
\bar{\gamma}_{V_{z},W}(r)=\frac{1}{4\pi r^{2}}\int\nu(W\cap V_{z,\boldsymbol{t}})d\boldsymbol{t}.
\]
The integration is carried out over the surface of a sphere of radius
$r$ in $\mathbb{R}^{3}$. This altered estimator is used since it
takes into account data that is in $W$ but not $V_{z}$ in the inner
summation. This reduces the truncation errors that could occur if
we used the estimator in Equation \eqref{eq:RadiallySymmPCF}. It is
in essence a generalization of our standard PCF estimator to allow
for the case that $\boldsymbol{r}_{i}$ and $\boldsymbol{r}_{j}$
correspond to different types of points (or points in different subdomains). 

Upon computation, we note that the mean square error\footnote{the mean square error is defined as $E=\int_{r_0}^{r_1}(\hat{g}_{V_{z}}(r)-\hat{g}(r))^2dr$. It is approximated by a trapezoid rule quadrature.} over $z$ remained
less than 0.05 for $r\in(0,2)$. We do observe some variation with
$z$ in the pair correlation function, however, it seems to be a minor
effect. In particular, we note that in some of the data sets, the
height of the first peak in $\hat{g}(r)$ decreases from the bottom
to the top of $W$. Although it would be ideal to compute a nonstationary
estimate for $g(\boldsymbol{r}_{1},\boldsymbol{r}_{2})$, in practice,
we found our calculations for such an estimator to be noisy, and unreliable
for small values of $\vert\boldsymbol{r}_{1}-\boldsymbol{r}_{2}\vert$
given the amount of data available. Thus, we have approximated the
variable pair correlation function with a SOIRS form of $\hat{g}(r)$
for use in our computations.

As a final note, alternative approaches for estimating $g(r)$ by
maximum likelihood estimation and Takacs-Fiksel estimation have been
explored in several papers \cite{baddeley2000practical,baddeley2000non,moller2003statistical,ripley1991statistical}.
With maximum likelihood estimation, the pair potential is assumed
to be a member of a predetermined class of functions that differ through
some set of parameters that can be optimized to fit the data. We did
not proceed with this approach due to the unusual structure of the
pair correlation function. With Takacs-Fiksel estimation, $g(r)$
is assumed to be a piecewise function (i.e. piecewise polynomial).
The weights of the coefficients in the piecewise approximant are found
by maximizing a nonlinear system of equations. Possible issues with
such an approach here are the computational cost to accurately resolve
$g(r)$ over a range of $r$ and the attainment of a continuous (not
just piecewise continuous) result. 

\subsection{Nearest Neighbor Distributions \label{subsec:Nearest-Neighbor-Distributions}}

Usually, direct numerical examination of the convergence in probability
of point processes is impractical due to limitations on the amount
of data available and the computational infeasibility of carrying
out tests needed to fully assess convergence. Although generally insufficient
to rigorously prove convergence, comparisons between summary statistics,
may be used to provide evidence of similarities between point processes. 

As will be discussed in subsequent sections, a statistical model designed
to match the first and second order statistical properties of experimental
data will be developed. Since lower order factorial product densities do not
fully characterize a point process, the assumption that the data
is well characterized by these statistics may be bolstered by
comparisons of summary statistics that depend on statistical correlations
of all orders. In this capacity, the $k$-nearest neighbor distributions
are useful. 

The $k$th-nearest neighbor distribution is defined as the probability
density function of distances between a point $\boldsymbol{r}\in\Phi$
and the $k$th closest point $\boldsymbol{r}^{\prime}\in\Phi\backslash\left\{ \boldsymbol{r}\right\} $.
Discussions about these distributions and their relation to moment
densities can be found in \cite{truskett1998density} and \cite{torquato2013random}.

Following \cite{torquato2013random,truskett1998density}, the \emph{nearest
neighbor distribution function}(NNDF), is defined as the probability,
\[
E_{P}^{(1)}(r)=\Pr\left[\min_{\boldsymbol{r}\in\Phi\backslash\{\boldsymbol{r}^{\prime}\}}\vert\boldsymbol{r}-\boldsymbol{r}^{\prime}\vert\leq r,\ \ \boldsymbol{r}^{\prime}\in\Phi\right].
\]
The \emph{nearest neighbor probability density}(NNPD)\emph{,} $H_{P}(r)$,
defined as the derivative of $E_{P}(r)$ with respect to $r$, is
often computed as well. The $k$th-nearest neighbor distributions
are defined by
\[
E_{P}^{(k)}(r)=\Pr\left[\min_{\{\boldsymbol{r}_{1},\dots\boldsymbol{r}_{k}\}\in\Phi\backslash\{\boldsymbol{r}^{\prime}\}}\left(\max_{i=1..k}\vert\boldsymbol{r}_{i}-\boldsymbol{r}^{\prime}\vert\right)\leq r,\ \ \boldsymbol{r}^{\prime}\in\Phi\right],
\]
\[
H_{P}^{(k)}(r)=\sum_{i=0}^{k-1}\left.\left(\frac{\partial}{\partial r^{\prime}}E_{P}^{(k)}(r^{\prime})\right)\right|_{r^{\prime}=r}.
\]
  Along with the intensity and pair correlation function, we use
the NNPDs as a means of comparing experimental data to realizations
of point processes generated through simulation. As shown in \cite{torquato2013random,truskett1998density},
$H_{P}^{(k)}(r)$ depends on relations of all orders between points
in the point processes, not just the lower order summary statistics
which can be estimated with the methods discussed in Section \ref{subsec:PairCorrelation}.
Although equivalence in NNPDs does not guarantee equivalence of two
point processes in probability, if two point processes are equal in
probability, they must have the same nearest neighbor distributions.
In practice, testing for equivalence in probability of a general point
process is computationally infeasible, and would require an inordinate
amount of data, however the nearest neighbor distributions are a useful
summary statistic due to their low dimensionality, and the ease with
which they can be estimated. 

Two disadvantages of nearest neighbor distributions as a means of 
comparing point processes are their lack of directional information,
 and that, as defined here, they are spatially homogeneous.
 Thus, the NNPDs we compute are best
understood as homogenized variants of a more general nearest neighbor
distribution that may depend on location. Since only four data sets
are available, the computation of a spatially variable NNPD would
be a formidable difficulty, and subject to greater variance than the
homogenized NNPD. 

\begin{figure}
\begin{centering}
\includegraphics[width=0.3\columnwidth]{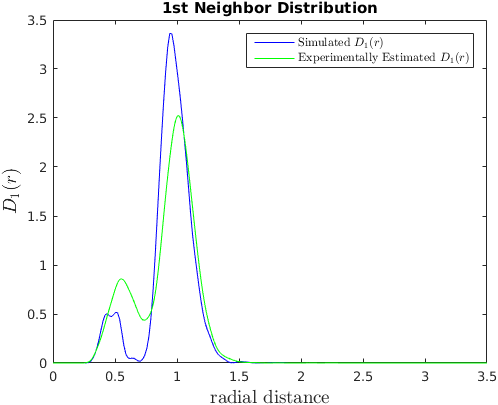}\quad{}\includegraphics[width=0.3\columnwidth]{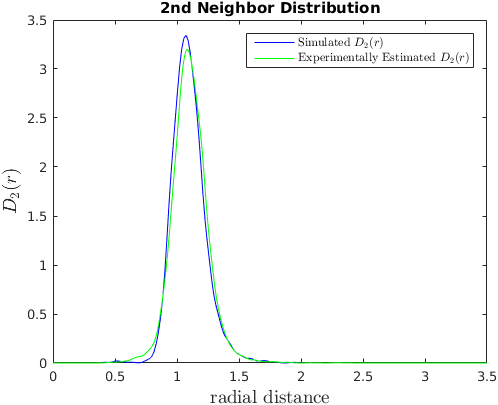}\quad{}\includegraphics[width=0.3\columnwidth]{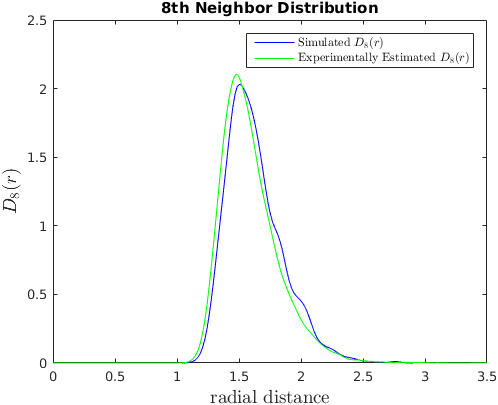}
\par\end{centering}
\caption{\label{fig:Comparison-of-nearest-neighbor}Comparison of nearest neighbor
density functions between a simulated biofilm, and the experimental
data. The bimodal behavior in the 1st neighbor distribution is not
completely captured by the model, however, higher NNDFs (2 and 8 depicted
here) are very accurately matched by the model.}
\end{figure}

\subsubsection{Nearest Neighbor Distribution Estimators}
To estimate the nearest neighbor distributions, we use \emph{minus
sampling} \cite{stoyanstochastic}. Minus sampling is the technique
of constructing estimators that ``leave out'' points near the edges
of the domain to mitigate edge effects. Minus sampling estimators
are less efficient than other estimators because they do not utilize all available data. However,
for statistics that are strongly influenced by edge effects, the benefit
in reducing edge effects can be well worth the inefficient use of
data. Using the symbol, $\ominus$, to denote Minkowski subtraction
\cite[\S 1]{stoyanstochastic}, the nearest neighbor distribution
is estimated as,
\[
\hat{H}_{P}(r)=\frac{1}{\Phi\left(W\ominus\mathcal{B}_{r_{c}}(0)\right)}\sum_{\boldsymbol{r}_{i}\in\Phi_{W}\ominus\mathcal{B}_{r_{c}}(0)}k_{b}\left(r-\left|\boldsymbol{r}_{i}-\underset{\boldsymbol{r}\in\Phi\cap W\backslash\{\boldsymbol{r}_{i}\}}{\arg\min}\vert\boldsymbol{r}_{i}-\boldsymbol{r}\vert\right|\right).
\]
Minus sampling is suitable in our application since the box-shaped
geometry of the domain dictates that most of the data is located away
from the edges. Thus, the loss of data due to minus sampling is not
severe. Even though there is a variation in number density near the top of the domain, we found that computations of the NNPDs that left out this portion of the data did not have a pronounced effect on the resulting estimate. 

Because $H_{P}(r)$ is a one dimensional probability density, kernel
density estimation methods of choosing $b$ based on the cardinality
of $\Phi_{W}$ are used \cite{parzen1962estimation,rosenblatt1956remarks}.
In particular, we use the \emph{ksdensity }function in Matlab to
compute $H_{P}(r)$ once $r_{c}$ is known. To specify $r_{c}$, we
first compute a preliminary estimate, $\hat{H}_{P,0}(r)$, without
minus sampling (e.g. assuming $r_{c}=0$), and then take $r_{c}$
equal to the supremum, 
\[
r_{c}=\underset{r}{\arg\sup}\left\{ \hat{H}_{P,0}(r)>0\right\} .
\]
This choice ensures that, with high probability, the nearest neighbor
of each point of $\Phi_{W}\ominus\mathcal{B}_{r_{c}}(0)$ is contained
in $\Phi_{W}$. Specifically, the expected error is the probability
$\Pr[\min_{\boldsymbol{r}_{j}\in\Phi}\vert\boldsymbol{r}_{i}-\boldsymbol{r}_{j}\vert>r_{c},\,\,\boldsymbol{r}_{i}\in\Phi\ominus\mathcal{B}_{r_{c}}(0)]$.
Since each data set contains $\sim4000$ bacteria locations, this
probability should be on the order of $10^{-3}$. A second bias term
arises from the use of non-parametric density estimators that have
finite support for nonzero $b$. We expect this source of bias to
be small since the asymptotic bias of kernel smoothing is typically
$\mathcal{O}(n^{-4/5})$ \cite{parzen1962estimation} where $n$ is
size of the data set ($\sim$4000 for each biofilm data set). For
$k>1$, analogous estimators are used with $r_{c}$ replaced by $r_{c}^{(k)}=\underset{r}{\arg\sup}\left\{ \hat{H}_{P,0}^{(k)}(r)>0\right\} $.
For larger values of $k$, the minus sampling technique is expected to become
inaccurate as the subset of $W$ from which points may be chosen shrinks.
However, we found that $r_{c}^{(k)}$ does not rapidly increase with
$k$, and at least for $k\leq20,$ the amount of data that must be disregarded due to minus sampling is small compared to the amount of data in each data set.
In Figure \ref{fig:Comparison-of-nearest-neighbor} we show the nearest
neighbor functions computed for the experimental data (averaged over
four data sets) and a realization of a statistical model discussed
in the next section. 

\section{Statistical Model for Bacterial Spatial Arrangement\label{sec:Replication-of-Biofilm}}

In this section, we introduce a statistical model for the spatial
arrangement of bacteria in biofilms, and discuss a method to generate
realizations of the model. Our approach, inspired by results from
the statistical physics of fluids \cite{hansen1990theory}, is to
characterize the first and second order interactions between bacteria
by the computation of empirical \emph{pair potential} and \emph{singlet
energ}y functions. These energy functions are related to the number
density and pair correlation functions computed in Section \ref{sec:Statistics-of-Biofilm}
through an integral equation known as the \emph{Ornstein-Zernike}
(OZ) equation, a closure relation which provides a formula for the
pair potential in terms of computable quantities, and a number density integral
equation, derived in \cite{lovett1976j} for the singlet potential. 

In our model, we apply a hypernetted-chain closure relation, and numerically
solve the OZ and the number density integral equation to obtain the pair
potential and singlet potential. Implicit in this approach is the
assumption 1st and 2nd order statistics are sufficient to reconstruct
the data. The resulting energy functions are then used to construct
probability density functions.

In practice, it is difficult to determine the normalizing constant
needed in the construction of the probability density function over configurations of points in the point process. To avoid
this issue, we use the energy functions obtained from the model to
construct unnormalized probability density functions. Realizations
of the model can then be generated through simulation using a Markov
chain Monte Carlo method (MCMC). The result of each simulation is
a set of points in a domain, $W$, which have similar statistical
properties, as measured by the tools discussed in Section \ref{sec:Statistics-of-Biofilm},
to the experimental data sets. 

\subsection{Pairwise Interaction Processes \label{subsec:Pairwise-Interaction-Processes}}

For a general Markov point process \cite[\S 6]{moller2003statistical}, the probability that a realization of the point process defined on a region, $W$, has some property, $F$,
can be written as
\[
\Pr(\Phi\in F)=\sum_{n=0}^{\infty}\frac{\exp(-\nu(W))}{n!}\underset{n}{\underbrace{\int_{W}\dots\int_{W}}}1[\Phi\in F]f_{n}(\boldsymbol{r}_{1},\boldsymbol{r}_2,\dots,\boldsymbol{r}_{n})d\boldsymbol{r}_{1}\dots d\boldsymbol{r}_{n}.
\]
Each term, $f_{n}(\boldsymbol{r}_{1},\dots,\boldsymbol{r}_{k})$ is
an $n$th order probability density function of the configuration of $n$ points. It is defined as the
conditional probability density given a realization of the point process
containing $n$ points, e.g. for an $n$-tuple of Borel sets, $\{B_1,\dots,B_n\}$,
\[
\mathbb{E}[\boldsymbol{x}_1\in B_1, \dots, \boldsymbol{x}_n\in B_n |\Phi(W)=n]
=\int_{B_1}\dots\int_{B_n} f^{(n)}(\boldsymbol{r}_1,\dots,\boldsymbol{r}_n)d\boldsymbol{r}_1\dots d\boldsymbol{r}_n.
\]
If there are only first and second order interactions between points,
the point process is known as a \emph{pairwise interaction process},
and the $n$th order probability can be written following \cite{hansen1990theory}
and \cite{moller2003statistical} as
\[
f_{n}(\boldsymbol{r}_{1},\dots\boldsymbol{r}_{n})=\frac{1}{Z_{n}}\exp\left[-\beta\sum_{i=1}^{n}\phi(\boldsymbol{r}_{i})-\beta\sum_{i=1}^{n}\sum_{j=i+1}^{n}v(\boldsymbol{r}_{i}-\boldsymbol{r}_{j})\right],
\]
where $\phi(\boldsymbol{r})$ is the \emph{singlet potential}, $v(\boldsymbol{r}_{1},\boldsymbol{r}_{2})$
is the\emph{ pair potential}, the normalizing constant, $Z_{k}$,
is known as a configurational integral (closely related to a thermodynamic
partition function) and, $\beta$ is inversely proportional to the
temperature. This is a vast simplification of the general Markov
point process as pairwise interaction processes are specified by just
two functions, $\beta\phi(\boldsymbol{r})$ and $\beta v(\boldsymbol{r}_{1},\boldsymbol{r}_{2})$.
Assuming that the point process governing the bacteria positions is
of this form, the next task is to estimate $\phi(\boldsymbol{r})$
and $v(\boldsymbol{r}_{1},\boldsymbol{r}_{2})$ from experimental
data.

\subsection{Computation of the Singlet and Pair Potentials\label{subsec:Computation-of-the-Potentials}}

The computation of the pair correlation function given a pair energy
function is a long standing and thoroughly studied problem in statistical
mechanics, especially in the case of a homogeneous system\cite{blum1976solution,kunkin1969inverse,ornstein1914influence}.
The key components needed to solve this problem are the solution of
the OZ equation, and the statement of a ``closure-relation'' that
relates the pair energy to computable thermodynamic functions. The
OZ equation, which can be derived through functional differentiation
of the thermodynamic grand cannonical ensemble, gives the definition
of a correlation function, known as the \emph{direct correlation function}
(DCF), in terms of the PCF and the number density. This is important
as the most successfully used closure relations are algebraic relations
between the DCF, PCF, and pair potential. In contrast to the problem
of using an assumed pair potential to obtain the PCF, here we discuss
the numerical solution of the inhomogeneous inverse problem: the computation
of a pair potential given a pair correlation function. Such inverse
solutions of the OZ equation appear to have first been studied in
\cite{kunkin1969inverse}, and numerical studies tend to favor Fourier
methods to solve the relevant equations \cite{2004fast}. 

For a variable number density, $\rho(\boldsymbol{r})$, the OZ equation
can be written as
\begin{equation}
h(\boldsymbol{r}_{1},\boldsymbol{r}_{2})=c(\boldsymbol{r}_{1},\boldsymbol{r}_{2})+\int\rho(\boldsymbol{r}_{3})c(\boldsymbol{r}_{1},\boldsymbol{r}_{3})h(\boldsymbol{r}_{2},\boldsymbol{r}_{3})d\boldsymbol{r}_{3}.\label{eq:OrnsteinZernike}
\end{equation}
The function, $h(\boldsymbol{r}_{1},\boldsymbol{r}_{2})\equiv g(\boldsymbol{r}_{1},\boldsymbol{r}_{2})-1$,
is known as the\emph{ indirect correlation function}, and $c(\boldsymbol{r}_{1},\boldsymbol{r}_{2})$
is the DCF. Equation (\ref{eq:OrnsteinZernike}) can be further simplified
if, allowing for arbitrary variation in the $z$-direction, we assume
that $h(\boldsymbol{r}_{1},\boldsymbol{r}_{2})=h(\vert\boldsymbol{r}_{1}-\boldsymbol{r}_{2}\vert)$
is translation invariant and isotropic, and that $c(\boldsymbol{r}_{1},\boldsymbol{r}_{2})$
is radially dependent in the $x$ and $y$ coordinates. The assumption
of a radially symmetric $h(r)$ is implied by the assumption of a
SOIRS process. The assumption of radial symmetry in the horizontal
plane is justified  since it matches the observed variation in number density.
It is also possible to show that, given a transversely isotropic pair
correlation function, and vertically variable number density, the
OZ equation admits transversely isotropic direct correlation functions.
Defining $\vert\boldsymbol{r}\vert_{xy}=\sqrt{x^{2}+y^{2}}$, and
using the assumption of transverse anisotropy, 
\begin{equation}
h(\vert\boldsymbol{r}_{1}-\boldsymbol{r}_{2}\vert)=c(\vert\boldsymbol{r}_{1}-\boldsymbol{r}_{2}\vert_{xy},z_{1},z_{2})+\int\rho(z_{3})c(\vert\boldsymbol{r}_{2}-\boldsymbol{r}_{3}\vert_{xy},z_{2},z_{3})h(\vert\boldsymbol{r}_{1}-\boldsymbol{r}_{3}\vert)d\boldsymbol{r}_{3}.\label{eq:AvgOZ}
\end{equation}
Equation (\ref{eq:AvgOZ}) can be written as a convolution over the
$xy$ plane, and an integration over the $z$ plane. With $\hat{\boldsymbol{e}}_{x}$
and $\hat{\boldsymbol{e}}_{y}$ representing the Cartesian unit vectors
in the $x$ and $y$ directions, setting $\boldsymbol{r}_{3}^{\prime}\cdot(\hat{\boldsymbol{e}}_{x}+\hat{\boldsymbol{e}}_{y})\leftarrow(\boldsymbol{r}_{3}-\boldsymbol{r}_{2})\cdot(\hat{\boldsymbol{e}}_{x}+\hat{\boldsymbol{e}}_{y})$,
we obtain
\[
h(\vert\boldsymbol{r}_{1}-\boldsymbol{r}_{2}\vert)=c(\vert\boldsymbol{r}_{1}-\boldsymbol{r}_{2}\vert_{xy},z_{1},z_{2})+\int\int\int\rho(z_{3})c(\vert(\boldsymbol{r}_{1}-\boldsymbol{r}_{2})-\boldsymbol{r}_{3}^{\prime}\vert_{xy},z_{1},z_{3})h(\vert\boldsymbol{r}_{3}^{\prime}\vert)dx_{3}^{\prime}dy_{3}^{\prime}dz_{3}.
\]

In order to determine the pair energy, a closure relation that defines
$v(r_{12},z_{1},z_{2})$ in terms of $g(|\boldsymbol{r}_{1}-\boldsymbol{r}_{2}\vert)$
and $c(r_{12},z_{1},z_{2})$ is required. We use the the hypernetted-chain
(HNC) equation, given as
\[
\beta v(r_{12},z_{1},z_{2})=\left(h(|\boldsymbol{r}_{1}-\boldsymbol{r}_{2}\vert)-c(r_{12},z_{1},z_{2})\right)-\log g(|\boldsymbol{r}_{1}-\boldsymbol{r}_{2}\vert).
\]
In addition to the HNC, a separate closure relation, known as the
Percus-Yevick (PY) closure is commonly applied. Both closure relations
are approximate, and, depending on the situation, one closure may
be more suitable than the other. In our case we chose to use the HNC
as we found our numerical results to be more stable as compared with
the PY equation, although for almost all values of $r_{12},z_{1},z_{2}$,
both the HNC and PY results were very similar. In Figure \ref{fig:The-pair-potential-heights},
the pair potential energy, obtained by applying the numerical methods
described below to the OZ equation with a HNC closure, is depicted. 

To complete the model, the singlet energy must be estimated. We use
an equation that relates the number density and DCF to an external (or singlet) energy. This relation was originally derived in \cite{lovett1976j} to describe density variation at a fluid-vapor interface, 
\begin{equation}
\beta\frac{\partial}{\partial z}\phi(z)=-\frac{\partial}{\partial z}\log\rho(z)+\int c(\boldsymbol{r}-\boldsymbol{r}^{\prime},z^{\prime},z)\frac{\partial}{\partial z^{\prime}}\rho(z^{\prime})d\boldsymbol{r}^{\prime}.\label{eq:ActivityEquation}
\end{equation}
Although derived for a different problem, we do not see anything in the derivation that makes the equations application to biofilm simulation invalid.
After calculating the energy derivative, integration is done to obtain
the singlet energy. Upon integration, we set the constant of integration
to 0 since only ratios of the form $e^{-\beta\phi(\boldsymbol{r}_{1})}/e^{-\beta\phi(\boldsymbol{r}_{2})}$
will be needed for simulations. 

Because the number density is known from empirical estimates, Equation (\ref{eq:AvgOZ})
can be solved independently of Equation (\ref{eq:ActivityEquation}).
Thus, our approach is to first solve Equation (\ref{eq:AvgOZ}) for
$c(r_{12},z_{1},z_{2})$, and then solve Equation (\ref{eq:ActivityEquation})
using $c(r_{12},z_{1},z_{2})$ to obtain $\phi(z)$. As we have previously
discussed the computation of $\rho(z)$ and $h(r_{12},z_{1},z_{2})$
(see Section \ref{sec:Statistics-of-Biofilm}), it remains to devise
a numerical solution method for Equations (\ref{eq:AvgOZ}) and (\ref{eq:ActivityEquation}). 

\subsection{Numerical Solution to the Singlet and Pair Potential Equations\label{subsec:Numerical-Solution-to}}

After obtaining estimates for $h(\vert\boldsymbol{r}_{1}-\boldsymbol{r}_{2}\vert)$
and $\rho(z)$ using the methods of Section \ref{sec:Statistics-of-Biofilm},
Equations (\ref{eq:AvgOZ}) and (\ref{eq:ActivityEquation}) must
be solved numerically. In the homogeneous case, the integral term
in the OZ equation becomes a convolution that can be efficiently and
accurately handled by Fourier transform methods. However, due to the
inhomogeneous number density, the integral over $z$ is not a convolution
in this present application. In fact, with a variable number density,
the OZ equation implies that the pair correlation and direct correlation
functions cannot both simultaneously be translation invariant. A proof
of this fact is straightforward and shown in the Appendix. Thus the
standard Fourier transform methods will not apply to our application.

As discussed in the previous section, to simplify Equation \ref{eq:OrnsteinZernike},
we assume that the $xy$ dependence of the pair correlation and direct
correlation functions is homogeneous with regard to within-plane translations
and is radially symmetric. This assumption makes the $xy$ integration
in Equation (\ref{eq:AvgOZ}) a two dimensional radially symmetric
convolution. The radially symmetric convolution of two functions can
be found through use of the two dimensional Hankel transform, denoted
$\mathcal{H}[\cdot]$, defined as the involutory transform
\[
F(k)=\mathcal{H}[f(r)]=2\pi\int_{0}^{\infty}J_{0}(2\pi kr)f(r)rdr
\]
where $J_{0}(kr)$ is the $0$th order Bessel $J$-function. The two
dimensional Hankel transform can be applied to the OZ equation with
$h(\boldsymbol{r}_{1}-\boldsymbol{r}_{2})$ and $r_{ij}=\vert(\boldsymbol{r}_{i}-\boldsymbol{r}_{j})\cdot(\boldsymbol{e}_{x}+\boldsymbol{e}_{y})\vert$
to obtain 
\[
\mathcal{H}[h](k,z_{1},z_{2})=\mathcal{H}[c](k,z_{1},z_{2})+\int\rho(z_{3})\mathcal{H}[h](k,z_{1},z_{2})\mathcal{H}[c](k,z_{1},z_{3})dz_{3}.
\]
In the discrete analog, we discretize $z$ and $r$ by setting $z_{m}=m\Delta z$,
$m=1,\dots,N$, and $r_{\ell}=j_{\ell}^{(0)}R/j_{N+1}^{(0)}$, where
$j_{\ell}^{(0)}$ is the $\ell$th root of the zeroth order Bessel
$J$ function, and $R$ is the maximum value at which the estimator,
$\hat{h}$ from Equation (\ref{eq:RadiallySymmPCF}) is computed.
We set $R=3\mu m$ since beyond this distance, the radially symmetric
pair correlation function showed little variation. A discrete analogue
of the Hankel transform is numerically computed in Matlab using the
algorithm devised in \cite{guizar2004computation} for each pair of
$z_{m}$ and $z_{n}$ with $m,n\in[1,N]$. 

Approximating the $z$-integral with the trapezoid method, the discrete
approximation, $C_{k_{\ell}z_{m}z_{n}}^{(N)}$ of $\mathcal{H}[c](k_{\ell},z_{m},z_{n})$
for each fixed value of $k_{\ell}$ and $z_{m}$, is found by solving
\begin{equation}
H_{k_{\ell},z_{m},z_{n}}^{(N)}=C_{k_{\ell},z_{m}z_{n}}^{(N)}+\sideset{}{^{\prime\prime}}\sum_{n^{\prime}=1}^{N}\hat{\rho}(n^{\prime}\Delta z)H_{k_{\ell},z_{n},z_{n^{\prime}}}^{(N)}C_{k_{\ell},z_{m},z_{n^{\prime}}}^{(N)}\Delta z.\label{eq:DiscreteHankel}
\end{equation}
The symbol $\sideset{}{^{\prime\prime}}\sum$ indicates that the first
and last term of the sum are halved. Equation (\ref{eq:DiscreteHankel})
can be written as a matrix equation, 
\begin{equation}
\boldsymbol{H}_{k_{\ell},z_{m}}^{(N)}=(\boldsymbol{\mathcal{I}}^{(N)}+\boldsymbol{\mathcal{M}}_{k_{\ell}}^{(N)})\boldsymbol{C}_{k_{\ell},z_{m}}^{(N)}\,\,\,\forall k_{\ell},\forall z_{m}\label{eq:DirectCorrelationSolve}
\end{equation}
where $\boldsymbol{\mathcal{I}}$ is the $N\times N$ identity matrix
and, 
\[
\mathcal{M}_{k,ij}^{(N)}=\hat{\rho}(i\Delta z)H_{k,z_{i}z_{j}}^{(N)}\Delta z.
\]
In principle, the matrix $\boldsymbol{\mathcal{I}}+\boldsymbol{\mathcal{M}}$
could be ill-conditioned. However, in the simulations we conducted,
the condition number of $\boldsymbol{\mathcal{I}}+\mathcal{\boldsymbol{M}}$
ranges only from 1 to 100. To verify that the error from matrix inversion
does not become excessive, we compute the norm of $(\boldsymbol{\mathcal{I}}+\boldsymbol{\mathcal{M}}_{k_{\ell}})^{-1}$
in each solve. For an error $\boldsymbol{e}$ in the estimation of
$\boldsymbol{h}$, the definition of the matrix norm ensures that
$\Vert(\boldsymbol{\mathcal{I}+}\boldsymbol{\mathcal{M}}_{k_{\ell}})^{-1}\boldsymbol{e}\Vert\leq\Vert(\boldsymbol{\mathcal{I}+}\boldsymbol{\mathcal{M}}_{k_{\ell}})^{-1}\Vert\Vert\boldsymbol{e}\Vert$.
Thus, as long as $\Vert(\boldsymbol{\mathcal{I}+}\boldsymbol{\mathcal{M}}_{k_{\ell}})^{-1}\Vert$
is not too large, the error will remain small. We also note that the
trapezoid rule approximation is second order accurate, although this
does not imply second order accuracy of the solution since it does
not account for errors in the estimation of $h$ by $\hat{h}$. We
observed that the condition number is largest for $\ell<5\Delta z$
($\mathcal{O}(10^{2})$ with the parameters we use), and decreases
rapidly with increasing $\ell$ to $\mathcal{O}(1)$. \footnote{The matrix norm used here is the 2-norm. This norm is also used in
the computation of the condition number.}

\begin{figure}
\begin{centering}
\subfloat[\label{fig:PCF_z491}]{\begin{centering}
\includegraphics[width=0.33\columnwidth]{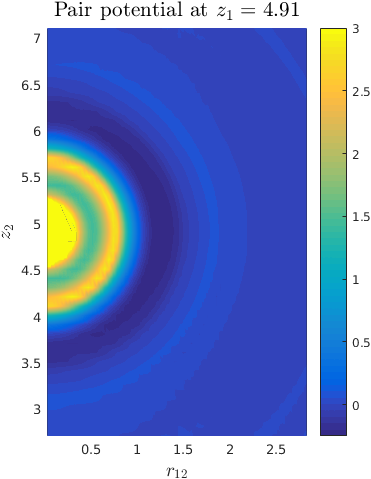}
\par\end{centering}
}\subfloat[\label{fig:PCF_z901}]{\begin{centering}
\includegraphics[width=0.33\columnwidth]{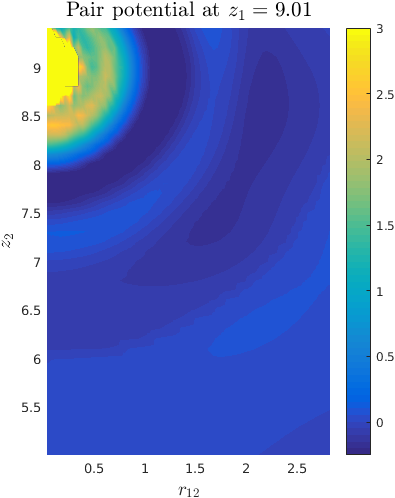}
\par\end{centering}
}
\par\end{centering}
\caption{\label{fig:The-pair-potential-heights}The potential, $v(r_{12},z_{1},z_{2})$
as it is computed from the hypernetted-chain equation. The two subfigures
show $v(r_{12},z_{1},z_{2})$ for two different values of $z_{1}$.
Although the pair potential at both values is similar, they are not
equivalent; the peaks and valleys in the pair correlation in \ref{fig:PCF_z901}
are slightly more pronounced than in \ref{fig:PCF_z491}. The differences
in the pair potential at different heights influences the height-dependent
number density trends we observed in the biofilm data.}
\end{figure}

Finally, after solving (\ref{eq:DirectCorrelationSolve}) for each
value of $\ell$ and $m$, the Hankel transform can be applied to
obtain $c_{r_{\ell},z_{m},z_{n}}^{(N)}$, the discrete approximation
of $c(r_{\ell},z_{m},z_{n})$. If there are $N_{r}$ radial nodes,
and $N_{z}$ vertical nodes, $N_{r}N_{z}$ equations must be solved,
and $2N_{z}^{2}$ transforms and inverse transforms must be computed.
Although this leads to poor scaling, since we only need to compute
the direct correlation function once, and can then store its value,
the cost is not prohibitive, and, for the values of $N_{r}$ and $N_{z}$
we use, can be found in under a minute. In fact, the computation of
$\hat{g}(r)$ is, by a substantial amount, the most time consuming
step, followed by the discrete Hankel transformations of $h^{(N)}$
and $c^{(N)}$ for each $z_{m}$, $z_{n}$ pair. The resulting pair
potential for data set \#3 is shown in Figure \ref{fig:The-pair-potential-heights}.

After computing $c^{(N)}$, a discrete approximation of the singlet
energy now can be obtained. Discretizing Equation (\ref{eq:ActivityEquation})
with the trapezoid rule, and using the estimate, $c_{r_{\ell},z_{m},z_{n}}^{(N)}$,
we arrive at an explicit approximation of $\phi^{\prime}(z)$, 
\[
\beta\phi_{z_{n}}^{\prime(N)}=-\frac{1}{\hat{\rho}(n\Delta z)}\hat{\rho}^{\prime}(n\Delta z)+2\pi\sideset{}{^{\prime\prime}}\sum_{m=1}^{N_{z}}\left(\sum_{\ell=1}^{N_{r}}c_{r_{\ell},z_{m},z_{n}}^{(N)}\Delta r_{\ell}\right)\hat{\rho}^{\prime}(m\Delta z)\Delta z.
\]
Given $\hat{\rho}^{\prime}(z)$, the discretization should be second
order accurate in $\Delta z$ and $\Delta r_{\ell}$. As with the
estimation of $c(r_{12},z_{1},z_{2})$, a rigorous error derivation
must take into account error terms due to approximations used in estimating
$\rho(z)$ by $\hat{\rho}(z)$ and $c(r_{12},z_{1},z_{2})$ by $c_{r_{\ell},z_{m},z_{n}}^{(N)}$.
For the approximation of $\phi(z)$, there is no advantage to using
Hankel transforms for the radial integral since the real-space computation
is explicit in $\phi(z)$. This differs from the pair energy computation
where the real space integral equation is a three dimensional integral
equation but, the transformed equation is a set of decoupled one dimensional
integral equations. In that case, transforming in the $r$ coordinate
is highly beneficial. We also note that other numerical quadrature
methods could be applied to compute the values of the integrals, however,
we expect that the dominant error source is from the approximation
of the pair correlation function and spatially variable number density.
Thus, we do not expect significant differences resulting from different
quadrature choices. 

\subsection{A Markov Chain Monte Carlo Algorithm \label{subsec:A-Markov-Chain-Algorithm}}

The pair energy function and singlet potential form the basis for
an MCMC algorithm to generate ``artificial'' biofilms. The practical
aspect of MCMC algorithms that makes them so useful is that they rely
only on unnormalized probability densities. Given two realizations
of a point process, denoted $\boldsymbol{X}_{1}$ and $\boldsymbol{X}_{2}$,
each containing $n$ points in $W$, and an unnormalized probability
density, $p^{(n)}(\boldsymbol{X}_{1})\propto f^{(n)}(\boldsymbol{X}_{1})$,
the ratio 
\begin{equation}
\frac{f^{(n)}(\boldsymbol{X}_{1})}{f^{(n)}(\boldsymbol{X}_{2})}=\frac{p^{(n)}(\boldsymbol{X}_{1})}{p^{(n)}(\boldsymbol{X}_{2})}=\frac{\exp\left[-\beta\sum_{i}\phi(\boldsymbol{r}_{i}^{(1)})-\beta\sum_{i<j}v(\boldsymbol{r}_{i}^{(1)},\boldsymbol{r}_{j}^{(1)})\right]}{\exp\left[-\beta\sum_{i}\phi(\boldsymbol{r}_{i}^{(2)})-\beta\sum_{i<j}v(\boldsymbol{r}_{i}^{(2)},\boldsymbol{r}_{j}^{(2)})\right]}\label{eq:MCMC_ratio}
\end{equation}
 can be computed efficiently in comparison to evaluating or approximating
the probability density, $f^{(n)}(\boldsymbol{X})$. If $\boldsymbol{X}_{2}$
differs from $\boldsymbol{X}_{1}$ by the location of only one point
such that $\boldsymbol{r}\in\boldsymbol{X}_{1}$ and $\boldsymbol{r}^{\prime}\in\boldsymbol{X}_{2}$,
then (\ref{eq:MCMC_ratio}) simplifies and can be computed in $\mathcal{O}(n)$
operations as 
\begin{equation}
\frac{f^{(n)}(\boldsymbol{X}_{1})}{f^{(n)}(\boldsymbol{X}_{2})}=\exp\left[-\beta(\phi(\boldsymbol{r}^{\prime})-\phi(\boldsymbol{r}))-\beta\sum_{i=1}^{n}v(\boldsymbol{r}_{i},\boldsymbol{r})+\beta\sum_{i=1}^{n}v(\boldsymbol{r}_{i},\boldsymbol{r}^{\prime})\right].\label{eq:MCMC_energyupdate}
\end{equation}
Because $\phi(\boldsymbol{r})$ and $v(\boldsymbol{r}_{1},\boldsymbol{r}_{2})$
are computed using the methods of Section \ref{subsec:Numerical-Solution-to}
on a grid and are not known analytically, interpolation is used to
approximate their values at the points, $\boldsymbol{r}_{i}\in\Phi$,
which are arbitrary points in space. Linear interpolation was the
chosen interpolation method because it is computationally inexpensive and accurate in our case since the grid of points on which the energy is computed on in the previous section has a small spatial step. Potential improvements could apply the ideas in \cite{conrad2016accelerating} obtain MCMC chains such that have reduced dependence on the choice of interpolation method. For boundary conditions,
we imposed periodicity in the two horizontal directions, and an impenetrable
boundary at the top and bottom of the domain.  Given $f^{(n)}$ (or
$p^{(n)}$), the MCMC algorithm proceeds with the steps detailed in
Algorithm \ref{alg:MCMC-algorithm}.

\begin{algorithm}
\begin{algorithmic}[1]
\State{Generate $n$ randomly placed points in a window, $W$. Denote this set of points $\boldsymbol{X}_0$.}
\State{Displace one point, $\boldsymbol{x}_i\in\boldsymbol{X}_k$, chosen at random, by a uniformly random displacement, $\delta\boldsymbol{x}$ and set 
\begin{equation*}\tilde{\boldsymbol{X}} = 
\left\{\left(\boldsymbol{X}_k\backslash\{\boldsymbol{x}_i\}\right)\cup\{\boldsymbol{x}_i+\delta\boldsymbol{x}\}\right\}.\end{equation*}}
\State{Compute
\begin{equation*} \alpha = \min\left(\frac{f^{(n)}(\tilde{\boldsymbol{X}})}{f^{(n)}(\boldsymbol{X}_k)},1\right)\end{equation*}
according to Equation \eqref{eq:MCMC_energyupdate} using interpolation as needed.}
\State{If $\tilde{\boldsymbol{X}}\subset W$, then with probability $\alpha$, set $\boldsymbol{X}_{k+1} = \tilde{\boldsymbol{X}}$. Otherwise repeat steps 2-4.}
\State{Set $k \leftarrow k+1$ and repeat steps 2-4 unless a \emph{convergence criterion} has been reached. If a convergence criterion has been reached, output $\boldsymbol{X}_{k+1}$ and exit. }
\end{algorithmic}

\caption{\label{alg:MCMC-algorithm}MCMC algorithm for generating realizations
of a point process with spatial characteristics similar to those of
the experimental data, conditioned on a known domain, $W$, of finite size. As a convergence criterion, we use the total
unnormalized energy, which we observed to level off after $\mathcal{O}(10^{5})$
update steps with an acceptance rate near 50\%. }
\end{algorithm}

In step 5 of Algorithm \ref{alg:MCMC-algorithm}, a halting criterion
must be used to determine whether to continue looping through steps
2-5 or to exit. Such criteria are usually based on easily computable
quantities such as the total unnormalized density, $p^{(n)}(\boldsymbol{X})$,
or, a characteristic such as the empirical pair correlation function
\cite[\S 8]{moller2003statistical}. In our case, we observed that,
with an update step drawn from a uniform distribution, $\delta\boldsymbol{X}\propto U([-1/2,1/2]^{3})$,
the total energy leveled off after $\mathcal{O}(10^{5})$ steps, and
the average number of acceptances was approximately 1/2 the number
of attempts. The computed value of $g(r)$ for each sample also stabilized
by this point in each case. Thus, as a halting criterion, we compute
the total energy of the sample every 10000 steps and stop after it
has leveled off. In practice, this lead to convergence within 500000
steps in each simulation that we conducted.

\begin{figure}
\begin{centering}
\includegraphics[width=0.45\columnwidth]{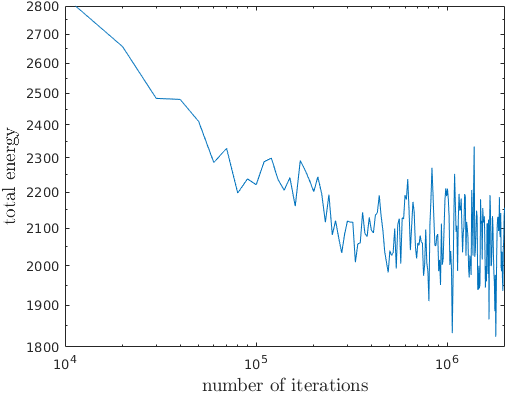}
\par\end{centering}
\caption{\label{fig:MCMC_Convergence}Convergence in the total energy, $E=\exp\left[-\beta\sum_{i}z(\boldsymbol{r}_{i})-\beta\sum_{i<j}v(\boldsymbol{r}_{i},\boldsymbol{r}_{j})\right]$
of the MCMC method described here. The result here is plotted on a
log-log scale to highlight the initial ``burn-in'' phase followed
by the attainment of an ``equilibrium'' energy. The energy is plotted
at every 10000th iteration. Even after the burn-in phases, there is
still variability in the energy between iterations, but there is no
downward trend in the energy. }
\end{figure}

The convergence in total energy of one particular realization is shown
in Figure \ref{fig:MCMC_Convergence}. There is a clearly distinguishable
initial period of decreasing energy followed by a valley where the
energy remains within a small range.  

The algorithm is conditional on $n$, however, for large values of
$n$, it has been observed that the difference between grand cannonical
and cannonical MCMC results is small \cite[\S 8]{moller2003statistical}.
If desired, it seems possible to augment the algorithm to include
addition and deletion steps or to allow $n$ to be a random variable
in step 1 of the algorithm to obtain an unconditional method. However,
such extensions also require knowledge of the distribution of $n$
over distinct samples. Since we only have four experimental data sets,
it seems unjustified to assume a distribution for $n$ without more
information. 

\section{Comparison of Material Properties\label{sec:Comparison-of-Material}}

Although experimental results on biofilms often range drastically
between different studies, it is generally agreed that over short
time scales and moderate mechanical stresses, biofilms behave as viscoelastic
materials \cite{pavlovsky2013situ}. One way to characterize biofilm rheology is through measurement of the dynamic moduli, which are defined in Table \ref{tab:Definitions-of-commonly}.

\begin{figure}
\begin{centering}
\subfloat[\label{fig:Gprime_comparison}]{\begin{centering}
\includegraphics[width=0.9\columnwidth]{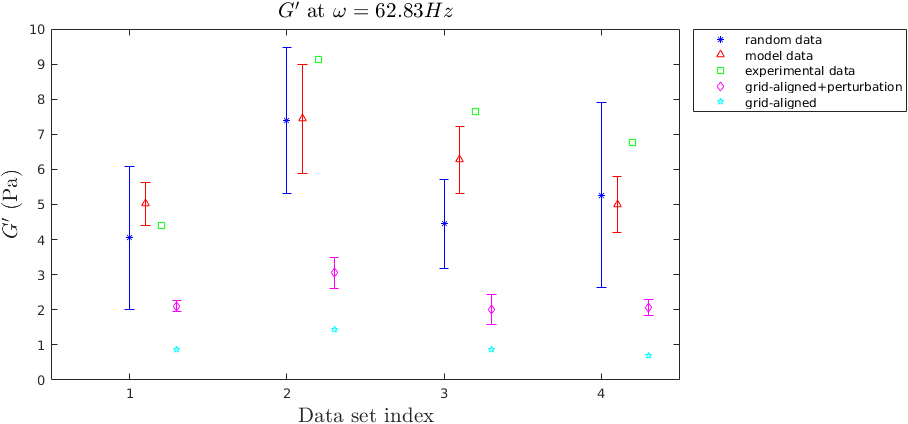}
\par\end{centering}
}\quad{}\subfloat[\label{fig:Gprime2_comparison}]{\begin{centering}
\includegraphics[width=0.9\columnwidth]{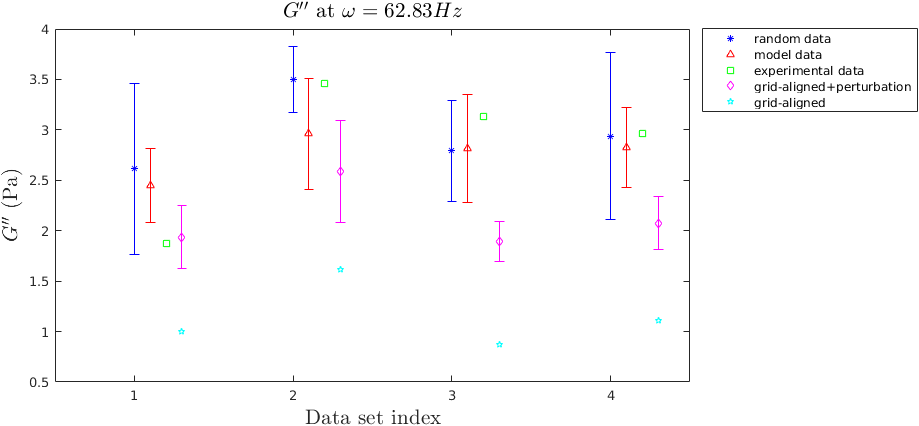}
\par\end{centering}
}
\par\end{centering}
\centering{}\caption{\label{fig:Comparison-of-Material} Figure \ref{fig:Gprime_comparison}
shows the comparison between the statistical models and the experimental data
for $G^{\prime}$ and Figure \ref{fig:Gprime2_comparison} shows the
the comparison for $G^{\prime\prime}$. Both figures are from simulations at $\omega=62.83Hz$.
For the grid aligned data, the data falls on regular grid of the form
$\boldsymbol{X}=(ih,jh,kh)$, for $i,j,k$ being integers. For the
Random Data and PCF Model, each entry corresponds to the average of
5 samples containing the same number of points and domain as each of four
experimental data sets. For the grid aligned plus perturbation, the
results are from five trials with the grid aligned data and a peturbation
drawn from a normal distribution of mean 0 and deviation of 0.2. The $x$ axis numbering coresponds to each of the four data sets the statistical models were designed to compare with. Error bars in Figure \ref{fig:Comparison-of-Material} are 95\% confidence intervals. These are computed using Matlab's \emph{normfit} function.}
\end{figure}

In Figure \ref{fig:Comparison-of-Material}, we depict the dynamic moduli,
for four different types of point processes and the experimental data. Since we have four experimental data sets available to us, we repeat the comparison for each data set. The ``random" data sets are realizations of a Poisson process of constant number density, equal to the average number density of the experimental data sets. The ``model" data is based on realizations of the model in Section \ref{sec:Replication-of-Biofilm}, the grid-aligned data use a regular grid of points in simulation, and the grid-aligned plus perturbation uses the same  grid-aligned data with an added perturbation drawn from a normal distribution of mean 0. 

We see that the uniformly random and PCF models perform better than a grid aligned
approximation, although some amount of improvement is seen when the
grid aligned data is randomly perturbed. In each case, the comparisons are made between point processes with approximately the same number of points as the experimental data. (In the uniformly random, and model based simulations, the number of points is exactly the same as the experimental data)

To estimate the variability in the material properties for each of the
point processes we have mentioned, the dynamic moduli are computed
over 5 trials. Although it would be beneficial to observe a larger
number of trials over a range of frequencies, the computational cost
of such an endeavor is currently prohibitive.

\section{Discussion\label{sec:Discussion}}

The results in Section \ref{sec:Comparison-of-Material} show how different models of the positions of bacteria in a biofilm can influence the mechanical properties of the simulated biofilm. 
It was found that the Poisson process model, and grid-aligned model do not
yield results that are consistent with statistical characteristics of the experimental data. However, we were surprised to find that despite this difference from the experimental data, the Poisson model does lead to biofilms with similar dynamic moduli, performing as well as our model in terms of recreating the mechanical properties of a biofilm. In contrast, the grid-aligned model, as shown in Figure \ref{fig:Comparison-of-Material} is not a close match. These findings indicate that nonuniformity can lead to stronger biofilms in comparison to the grid-aligned case. We also show that the model introduced in Section \ref{sec:Replication-of-Biofilm}, with first and second
order characteristics informed by experimental data, yields agreement in the material properties and agreement in statistical properties of the data.

Although the uniformly random model exhibits similar dynamic moduli as experimental data, from a physical interpretation, it is clear that the bacteria positions
cannot conform to a Poisson process. The lack of correlations between point locations, a defining feature of Poisson processes, implies
that for any radius, $r$, there is a nonzero probability that within
a given realization, two points of the process are separated by a distance
less than $r$. In contrast, bacteria have finite radii and are impenetrable,
thus the centers of mass of two bacteria cannot be separated by less
than some hard-sphere radius. The hard-sphere property of the experimental
data is readily apparent upon computation of the nearest neighbor
distribution and the pair correlation function. 


In contrast to our observations, we note that in
\cite{wrobel2014modeling}, the network topology of an immersed boundary
method model was not observed to have strong effects on bulk viscoelastic
properties of the system. However, we believe that this difference
is due to the magnitude of the number density of the point process being simulated and the average connectivity of each point. We choose a connectivity
model with an average connectivity of close to 9 connections per bacteria, whereas
\cite{wrobel2014modeling} used a regular network of points with 27
linkages per node. It is our conjecture that at high number densities, and high levels of connectivity, the spatial positioning has less of an impact on rheological features, whereas for lower density and lower connectivity situations, the spatial positioning has a stronger effect. In a future work we plan to further study this idea.

In this paper, the focus was solely on the spatial arrangement of
the centers of mass of bacteria. No attention was given to developing
the ``network'' or clustering statistics of bacteria, i.e. what
is the chance that two bacteria whose separation is $\boldsymbol{r}$
are connected by viscoelastic links, and how strong is such a link
expected to be? The reason for this omission is that current state
of the art experiments allow for determination of bacteria locations,
but to the extent of our knowledge, the connectivity of bacteria in a biofilm has not been experimentally measured. Thus, we use a simple distance based
connectivity rule, assuming that bacteria within a certain radius
are connected by a spring which cannot bend, as was originally proposed
in \cite{alpkvist2007description}, and used again in \cite{hammond2013viscosity}
and \cite{stotsky2015rheology}. Although this model is simple it
yields realistic results when combined with appropriate viscoelastic
models of the linkages. However, a more realistic connectivity model
might impact the observed material properties. It would also be of interest to determine a mechanism by which springs can be ruptured due to the accumulation of stress such as in \cite{sudarsan2015simulating}.

Although not the focus of our work, an interesting result that
arose in the course of this study was the observation of oscillations
in the number density of bacteria near the fluid-biofilm interface.
It would be interesting to discover whether this variation in number density
is a passive effect due to fluid motion at the biofilm-fluid surface
or other experimental conditions, or whether it is a strategy employed
by biofilm forming bacteria to improve the biological fitness of a
given biofilm. 

\section{Acknowledgments}
\label{sec:Acknowledgments}

This work was supported in part by the National Science Foundation
grants PHY-0940991 and DMS-1225878 to DMB, and by the Department of
Energy through the Computational Science Graduate Fellowship program,
DE-FG02-97ER25308, to JAS. The authors would also like to thank Mike Solomon (University of Michigan) and John Younger (Akadeum Life Sciences) for insightful discussions and suggestions concerning this work.

\bibliographystyle{EJAM}
\bibliography{BibtexFile}

\appendix

\section{}

\subsection{\label{subsec:Estimation-Density}Bandwidth Selection for the Estimation
of the Number Density}

In Section \ref{sec:Replication-of-Biofilm}, it was necessary to
estimate the number density of biofilms samples. In order complete
this task, we used a kernel density based estimator of the form
\[
\hat{\rho}(z;b)=\frac{1}{A}\sum_{\boldsymbol{r}_{i}\in\Phi\cap W}\frac{k_{b}(z-\hat{\boldsymbol{e}}_{z}\cdot\boldsymbol{r}_{i})}{c(z;b)}.
\]
As is typical with kernel density estimation, a choice of the scale
parameter, $b$, must be made. Several typical techniques are discussed
in \cite{wand1993comparison}. The general idea is to minimize the
\emph{mean integrated square error} (MISE), 
\[
MISE(\hat{\rho}(z;b))=\int\left(\hat{\rho}(z;b)-\rho(z)\right)^{2}dz.
\]
The difficulty is of course the lack of knowledge of $\rho(z)$. Although
the choice of $b$ is partly intuitive, (e.g. too large a value leads
to an overly smooth estimate, and too small a value leads to an overly
jagged estimate), it is difficult to judge the best value among reasonable
values of $b$ by mere qualitative observation. Although there are
numerous methods of bandwidth selection, we choose to use the Least Squares Cross-Validation (LSCV) method
described in \cite{guan2008consistent} as a first estimate. We also found that
by visual examination values of $b$ in the range $(0.13,\,0.3)$
seem to provide reasonable results, and thus expect any optimization
method to yield a value in this range. From \cite{guan2008consistent},
we optimize, 
\[
LSCV(b)=\int_{W}\hat{\rho}(z;b)^{2}dz-2\sum_{\boldsymbol{r}_{i}\in\Phi\cap W}\hat{\rho}(z_{i};b)-k(0;b)/(Ac(z_{i};b))
\]
\[
=\frac{1}{A^{2}}\sum_{\boldsymbol{r}_{i}\in\Phi\cap W}\sum_{\boldsymbol{r}_{j}\in\Phi\cap W}\int_{W}\frac{k_{h}(z-z_{i})k_{h}(z-z_{j})}{c_{b}(z)^{2}}dz-\frac{2}{A}\sum_{\boldsymbol{r}_{i}\in\Phi\cap W}\sum_{\boldsymbol{r}_{j}\in\Phi\cap W\backslash\boldsymbol{r}_{i}}\frac{k_{b}(z_{i}-z_{j})}{c_{b}(z_{i})}.
\]
With the Epanechnikov kernel, and ignoring edge effects for simplicity,
\[
I(z_{i};b)\equiv\int_{W}k_{b}(z-z_{i})k_{b}(z)dz=\begin{cases}
\begin{array}{c}
\frac{3}{160b}\left(32-40(z_{i}/b)^{2}+20|z_{i}/b|^{3}-|z_{i}/b|^{5}\right)\\
0
\end{array} & \begin{array}{c}
|z_{i}|\leq2b\\
|z_{i}|>2b
\end{array}\end{cases}
\]
\[
LSCV(b)=\frac{1}{A}\left(\frac{3}{5b}\Phi(W)+\sum_{\boldsymbol{r}_{i}\neq\boldsymbol{r}_{j}}I(z_{i}-z_{j};b)\right)-\frac{2}{A}\sum_{\boldsymbol{r}_{i}\neq\boldsymbol{r}_{j}}k_{b}(z_{i}-z_{j})
\]

\begin{figure}
\begin{centering}
\includegraphics[width=0.4\columnwidth]{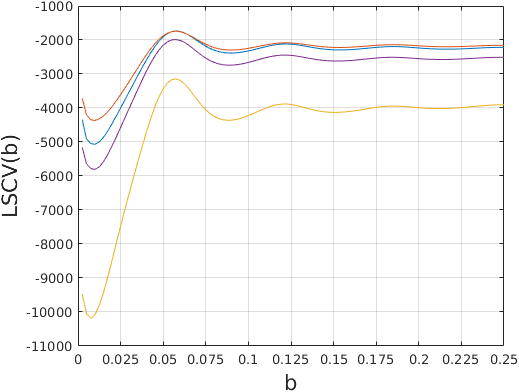}
\par\end{centering}
\caption{\label{fig:The-Least-Squares-CV}The Least Squares Cross Validation
value for selected values of $b$ over four data sets. }
\end{figure}

A plot of $LSCV(b)$ versus $b$ is shown in Figure \ref{fig:The-Least-Squares-CV}.
It can be seen that there exists several minima in each case, and
the question is how to choose the ``best'' minima. In each case,
the first minima, which is also the global minimum, is clearly too
small leading to density estimates with unacceptably high variance. We found that the shallow
local minimum at $b\approx0.21$ worked best in practice. We additionally implemented the log-likelihood
estimator described in \cite{cronie2016bandwidth} and found very
similar results. It is unclear if there exists a method that possess
a unique minimum in our case. There is also some evidence that spurious
local minimizers of LSCV functionals tend to occur at smaller values
than the optimal $b$, thus choosing the largest local minimizer seems
a suitable strategy \cite{hall1991local}.

\subsection{\label{subsec:Estimation-PCF}Bandwidth Selection for Estimation
of the Pair Correlation Function}

Similar to the estimation of a number density, the estimation of a
pair correlation function bandwidth has an important effect upon the
resulting estimator. From a theoretical perspective, it is easiest
to analyze estimators of the form 
\[
\hat{g}(r)=\frac{1}{4\pi r^{2}\bar{\gamma}_{W}(r)\rho^{2}}\sum_{\boldsymbol{r}_{i}\in\Phi\cap W}\sum_{\boldsymbol{r}_{j}\in\Phi\cap W\backslash\{\boldsymbol{r}_{i}\}}k_{b}(r-\vert\boldsymbol{r}_{i}-\boldsymbol{r}_{j}\vert)
\]
where $\bar{\gamma}_{W}(r)$ is known as the \emph{isotropized set
covariance \cite{stoyanstochastic,stoyan_estimation_1993}}. It can
be computed as the following integral over $\boldsymbol{t}\in\left\{ \boldsymbol{s}\in\mathbb{R}^{3}|\,\,\vert\boldsymbol{s}\vert=r\right\} ,$
\[
\bar{\gamma}_{W}(r)=\frac{1}{4\pi r^{2}}\int\nu(W\cap W_{\boldsymbol{t}})d\boldsymbol{t}.
\]
 Although generally intractable to integrate analytically, for boxes,
the integral can be computed:
\begin{align*}
\bar{\gamma}_{W}(r)= & \text{\ensuremath{\frac{1}{4\pi}\int\int\nu}(W\ensuremath{\cap W_{\boldsymbol{t}=(r,\theta,\phi)}})\ensuremath{\sin\theta}d\ensuremath{\phi}d\ensuremath{\theta}}\\
= & \frac{2}{\pi}\int_{0}^{\pi/2}\int_{0}^{\pi/2}(l-r\sin\phi\cos\theta)(w-r\sin\phi\sin\theta)(h-r\cos\phi)\sin\theta\,d\theta d\phi\\
= & hlw-\frac{1}{2}(hl+lw+wh)r+\frac{2}{3\pi}(h+l+w)r^{2}-\frac{1}{4\pi}r^{3}
\end{align*}
for $r<\min(h,l,w)$. Then, following \cite{ripley1991statistical}
and \cite{stoyan_estimation_1993}
\[
\mbox{Var}(\hat{g}(r))\approx\frac{0.6g(r)}{4\pi b\rho^{2}r^{2}\bar{\gamma}_{W}(r)}.
\]
It is noted in \cite[\S 5]{stoyan_estimation_1993} that this approximation
is particularly accurate for hard-core processes. The bias can be
computed as
\[
\mbox{Bias}(\hat{g}(r))=\left(\int k_{b}(r-r^{\prime})g(r^{\prime})dr^{\prime}-g(r)\right)^{2}
\]
Since the pair correlation function may exhibit a jump discontinuity
near the hard-sphere radius, it is helpful to examine the integrated
bias 
\[
\int\mbox{Bias}(\hat{g}(r))dr=\int\left(\int k_{b}(r-r^{\prime})g(r^{\prime})dr^{\prime}-g(r)\right)^{2}dr
\]
Where $g(r)$ is twice differentiable, the bias and variance can be
combined to obtain an approximation for the mean square error as a
function of $r$ and $b$,
\[
\mathbb{E}\left[\left(g(r)-\hat{g}(r;b)\right)^{2}\right]=\frac{0.6g(r)}{4\pi b\rho^{2}r^{2}\bar{\gamma}_{W}(r)}+\int k_{b}(r-r^{\prime})r^{\prime2}dr^{\prime}g^{\prime\prime}(r).
\]
Assuming $r>b$ since the hard-sphere diameter of bacteria is fairly
large in comparison to the range over which we compute $g(r)$, 
\begin{equation}
\mathbb{E}\left[\left(g(r)-\hat{g}(r;b)\right)^{2}\right]=\frac{0.6g(r)}{4\pi b\rho^{2}r^{2}\bar{\gamma}_{W}(r)}+\frac{1}{5}(b^{2}+r^{2})g^{\prime\prime}(r).\label{eq:MSE}
\end{equation}
Equation (\ref{eq:MSE}) can be minimized for each value of $r$ to
yield an analogous result to those typically derived in the case of
probability density estimators (c.f. \cite{parzen1962estimation,rosenblatt1956remarks}),
\[
b(r)=\left(\frac{3}{8}\frac{g(r)}{\rho^{2}r^{2}\bar{\gamma}_{W}(r)g^{\prime\prime}(r)}\right)^{1/3}.
\]
As is the case with density estimation, this expression is of limited
usefulness since it depends on the unknown quantities, $g(r)$ and
$g^{\prime\prime}(r)$. However, bandwidth selection methods, such
as \emph{least squares cross validation} (LSCV), and \emph{biased
cross validation} (BCV) \cite{guan2008consistent,wand1993comparison}
can be applied to the integrated MSE to estimate an optimal value
of $b$ across the entire interval. For instance, the LSCV estimator
for $g(r)$ can be formulated as in \cite[Equation 4]{guan2007least}.
In addition, to the minimization method introduced in \cite{guan2007least},
we also employ a ``binning'' technique to estimate optimal values
of $b$ over disjoint portions of the overall interval of computation.
The motivation for this adaptation is that the behavior of $g(r)$
is quite different near the hard-sphere radius in comparison to the
asymptotic behavior for $g(r)$ as $r$ grows. It seems sensible that
different scale parameters should be applied in these different regimes.
Thus, we employ LSCV functionals of the form 
\begin{equation}
LSCV(h;[r_{0},r_{1}])=4\pi\int_{r_{0}}^{r_{1}}\hat{g}(r;b)^{2}r^{2}dr-2\sideset{}{^{\neq}}\sum_{r_{0}\leq\vert\boldsymbol{r}_{i}-\boldsymbol{r}_{j}\vert\leq r_{1}}\frac{\hat{g}^{-(\boldsymbol{r}_{i},\boldsymbol{r}_{j})}(\vert\boldsymbol{r}_{i}-\boldsymbol{r}_{j}\vert;h)}{\bar{\gamma}_{W}(\vert\boldsymbol{r}_{i}-\boldsymbol{r}_{j}\vert)\rho(\boldsymbol{r}_{i})\rho(\boldsymbol{r}_{j})}.\label{eq:LSCV_PCF}
\end{equation}
The symbol $\hat{g}^{-(\boldsymbol{r}_{1},\boldsymbol{r}_{j})}(r;h)$
indicates the computation of the pair correlation function with points
$\boldsymbol{r}_{i}$ and $\boldsymbol{r}_{j}$ ignored. The expectation
of the summation in Equation (\ref{eq:LSCV_PCF}) is shown in \cite{guan2007least}
to converge in the limit of a large domain to 
\[
\int\hat{g}(r;h)g(r;h)4\pi r^{2}dr.
\]
One can see from this the similarity to classical LSCV estimation
of scale parameters for kernel density estimation \cite{wand1993comparison}.

In practice, we found that the summation of $\hat{g}^{-(\boldsymbol{r}_{1},\boldsymbol{r}_{2})}(\boldsymbol{r}_{1},\boldsymbol{r}_{2})$
was too expensive, leading to extremely lengthy computations. To alleviate
this issue, we approximated $\hat{g}^{-(\boldsymbol{r}_{1},\boldsymbol{r}_{2})}(\boldsymbol{r}_{1},\boldsymbol{r}_{2})$
as 
\[
\hat{g}^{-(\boldsymbol{r}_{1},\boldsymbol{r}_{2})}(\vert\boldsymbol{r}_{12}\vert)\approx\mathcal{I}\hat{g}(\vert\boldsymbol{r}_{12}\vert)-\frac{2}{4\pi\vert\boldsymbol{r}_{12}\vert^{2}\bar{\gamma}(\vert\boldsymbol{r}_{12}\vert)}\sum_{\boldsymbol{r}_{i}\neq\{\boldsymbol{r}_{1},\boldsymbol{r}_{2}\}}\left(k_{b}(\vert\boldsymbol{r}_{12}\vert-\vert\boldsymbol{r}_{1i}\vert)+k_{b}(\vert\boldsymbol{r}_{12}\vert-\vert\boldsymbol{r}_{2i}\vert)\right)
\]
where $\mathcal{I}\hat{g}(\cdot)$ is the linear interpolant of $\hat{g}(\cdot)$
to some value (i.e. $\vert\boldsymbol{r}_{12}\vert$). The factor
of 2 in the numerator of the second term is due arises since the terms
involving $\boldsymbol{r}_{1i}$,$\boldsymbol{r}_{2i}$, $\boldsymbol{r}_{i1}$,
and $\boldsymbol{r}_{i2}$ must be subtracted from $\hat{g}(r)$.
However, since $\vert\boldsymbol{r}_{ji}\vert=\vert\boldsymbol{r}_{ij}\vert$,
there are only $\Phi(W)$ unique terms in the sum. With this fix,
once $\hat{g}(r)$ is known, $\hat{g}^{-(\boldsymbol{r}_{1},\boldsymbol{r}_{2})}(\vert\boldsymbol{r}_{12}\vert)$
can be approximated in $\mathcal{O}(\Phi(W))$ computations as opposed
to $\mathcal{O}(\Phi(W){}^{2})$ computations. As long as $\hat{g}(\cdot)$
is initially computed on a sufficiently dense set of points, the interpolant
will be quite accurate. The resulting optimal values of $b$ for different
ranges of $r$ are shown in Table \ref{tab:Values-of-b}. In practice,
the value of $b(r)$ is assumed to be piecewise linear in $r$ taking
on the reported value in \ref{tab:Values-of-b} at the midpoint of
each interval. This prevents artificial discontinuous at the end points
of the intervals over which $b$ was estimated. Of course, if the
estimator is accurate, one would expect such discontinuities to be
small. Indeed, when $b(r)$ is a piecewise constant, discontinuities
are difficult to discern by sight. We also use numerical integration
to compute the relevant integrals in Equation (\ref{eq:LSCV_PCF}).

\begin{table}
\caption{\label{tab:Values-of-b}Values of $b$ determined through LSCV optimization
are reported for intervals of $r$. }
\begin{centering}
\begin{tabular}{c c c}
\hline 
$b$ & $r_{min}$ & $r_{max}$\tabularnewline
\hline 
\hline 
0.050 & 0.3 & 0.7\tabularnewline
\hline 
0.3700 & 0.7  & 1.6\tabularnewline
\hline 
0.2150 & 1.6  & 2.5\tabularnewline
\hline 
0.3960 & 2.5  & 5.0\tabularnewline
\hline 
\end{tabular}
\par\end{centering}

\end{table}

One final issue with LSCV bandwidth selection is the presence of multiple
minima. For the pair correlation function LSCV functional, spurious
minima near $b=0$ were observed. For the probability density estimation,
it has been suggested that spurious minima are usually less than the
ideal bandwidth \cite{hall1991local}. Thus, when multiple minima
are present, we choose the minima that is largest over the range of
values we consider for $b$. In Figure \ref{fig:Variation-in-LSCV},
we show the value of $LSCV(b;[0.6,0.8])$ as $b$ is varied.

\begin{figure}
\caption{\label{fig:Variation-in-LSCV}Variation in $LSCV(b;[0.6,0.8])$ as
$b$ changes from $0.001$ to 0.30. The presence of a spurious minimum
near $b=0$ is present, followed by a minimum near $b=0.08$. The
pair correlation function estimated over this interval for the two
minima are shown. The pair correlation function estimate corresponding
to the smaller value of $b$ exhibits marked oscillations which we
believe to be an indication of undersmoothing. In contrast, the larger
minima yields a relatively smooth curve. }
\end{figure}

An alternative approach to variable scale parameter estimation is
use techniques such as that introduced in \cite{abramson1982bandwidth}.
We believe such an approach would be effective as well.

\subsection{\label{subsec:Inhomogeneity-of-OZ}Inhomogeneity of the Direct Pair
Correlation Function in Nonstationary Processes}

The motivation for using a transversely anisotropic pair correlation
and direct correlation functions was attributed to properties of the
Ornstein-Zernike equation. In this section we demonstrate why the
Ornstein-Zernike equation implies a loss of translation invariance
in the pair correlation function and direct correlatioin function
when the density is variable. 

Consider the inhomogeneous Ornstein-Zernike equation as shown in Equation
(\ref{eq:OrnsteinZernike}). Let the two pairs of points $\{\boldsymbol{r}_{1},\boldsymbol{r}_{2}\}$
and $\{\boldsymbol{r}_{1}^{\prime},\boldsymbol{r}_{2}^{\prime}\}$
satisfy $\boldsymbol{r}_{1}-\boldsymbol{r}_{2}=\boldsymbol{r}_{1}^{\prime}-\boldsymbol{r}_{2}^{\prime}=\delta\boldsymbol{r}$,
and assume that the pair correlation and direct correlation functions
are translation invariant. Then, using the transformation $\boldsymbol{r}_{1}^{\prime}=\boldsymbol{r}_{1}+\boldsymbol{x}$,
$\boldsymbol{r}_{2}^{\prime}=\boldsymbol{r}_{2}+\boldsymbol{x}$,
\begin{align*}
\,\,\,\,\,h(\boldsymbol{r}_{1}-\boldsymbol{r}_{2})= & c(\boldsymbol{r}_{1}-\boldsymbol{r}_{2})+\int\rho(\boldsymbol{r}_{3})c(\boldsymbol{r}_{1}-\boldsymbol{r}_{3})h(\boldsymbol{r}_{2}-\boldsymbol{r}_{3})d\boldsymbol{r}_{3}\\
-\,\,h(\boldsymbol{r}_{1}^{\prime}-\boldsymbol{r}_{2}^{\prime})= & c(\boldsymbol{r}_{1}^{\prime}-\boldsymbol{r}_{2}^{\prime})+\int\rho(\boldsymbol{r}_{3})c(\boldsymbol{r}_{1}^{\prime}-\boldsymbol{r}_{3})h(\boldsymbol{r}_{2}^{\prime}-\boldsymbol{r}_{3})d\boldsymbol{r}_{3}\\
\hline 0= & \int\rho(\boldsymbol{r}_{3})c(\boldsymbol{r}_{1}-\boldsymbol{r}_{3})h(\boldsymbol{r}_{2}-\boldsymbol{r}_{3})d\boldsymbol{r}_{3}-\int\rho(\boldsymbol{r}_{3})c(\boldsymbol{r}_{1}^{\prime}-\boldsymbol{r}_{3})h(\boldsymbol{r}_{2}^{\prime}-\boldsymbol{r}_{3})d\boldsymbol{r}_{3}
\end{align*}
In order for this to be true for all translations, $\boldsymbol{x}$,
it must be the case that $\rho(\boldsymbol{r}_{3})=const$ almost
everywhere. Thus, for a smoothly varying number density, it cannot be the
case that $\boldsymbol{c}(\boldsymbol{r}_{1},\boldsymbol{r}_{2})$
and $h(\boldsymbol{r}_{1},\boldsymbol{r}_{2})$ are both simultaneously
translation invariant. 
\end{document}